\documentclass[a4paper,11pt]{article}
\usepackage{amsmath,amssymb,bm,epsfig,color,cite,braket,mathrsfs,slashed,multirow,graphicx,latexsym,cancel,comment}

\renewcommand{\thefootnote}{\fnsymbol{footnote}}

\newcommand{\D}{{\rm d}}

\newcommand{\CNB}{{\rm CNB}}

\newcommand{\Tritium}{{}^{3}{\rm H}}
\newcommand{\He}{{}^{3}{\rm He}}
\newcommand{\Eref}[1]{Eq.~\eqref{#1}}
\newcommand{\Erefs}[1]{Eqs.~\eqref{#1}}

\usepackage{subcaption}

\usepackage{hyperref,xcolor}
\hypersetup{
colorlinks=true,
linktocpage=true,
linkcolor=blue,
filecolor=blue,      
urlcolor=blue,
citecolor=blue,
   }
    
\numberwithin{equation}{section}

\begin{document}

\title{
{\Large \bf Pathways and impediments \\ towards a detection of the relic neutrino wind\\*[15pt]}
}

\author{
Andrew J.~Long$^{1}$,\footnote{
\href{mailto:andrewjlong@rice.edu}{andrewjlong@rice.edu}}\ \ 
Michiru Uwabo-Niibo$^{2}$,\footnote{
\href{mailto:michiru@ibs.re.kr}{michiru@ibs.re.kr}}\ \ 
Masahide Yamaguchi$^{2,3,4}$\footnote{
\href{mailto:gucci@ibs.re.kr}{gucci@ibs.re.kr}}
\\*[15pt]
$^{1}${\it\small
Department of Physics and Astronomy, Rice University, Houston, Texas 77005, U.S.A.} 
\\*[5pt]
$^{2}${\it\small
Cosmology, Gravity, and Astroparticle Physics Group, Center for
Theoretical Physics of the Universe, }\\{\it\small  Institute for Basic Science (IBS), Daejeon,
34126, Korea} 
\\*[5pt]
$^{3}${\it \small
Department of Physics, Institute of Science Tokyo,}\\{\it \small
2-12-1 Ookayama, Meguro-ku, Tokyo 152-8551, Japan}
\\
$^{4}${\it \small
Department of Physics and IPAP,
Yonsei University, 50 Yonsei-ro, Seodaemun-gu, Seoul 03722, Korea}
\\*[50pt]
}

\date{
\centerline{\small \bf Abstract}
\begin{minipage}{0.9\linewidth}
\medskip \medskip \small 
A direct detection of the cosmic neutrino background (CNB) in laboratories on Earth has been called the ``holy grail'' of experimental neutrino physics, but a still more glorious prize awaits.  
Beyond simply detecting the presence of relic neutrinos and measuring their flux, one may aspire to measure their energy distribution, polarization, anisotropies, temporal variation, and other properties.  
In this work we focus on the CNB wind, which is the approximately dipolar anisotropy in the CNB flux resulting from the relative velocity of the CNB rest frame and the lab frame.  
We consider a CNB detection strategy based on measuring the angular distribution of recoiling electrons at the tritium $\beta$-decay endpoint.  
In order to quantify the difficulty of detecting the CNB wind, we calculate the required exposure (detector mass times observation duration) for a $3\sigma$ discovery.
We find that detecting the CNB wind would require an exposure that is at least $10^{5}$ times larger than what's required for detecting the CNB flux alone.  
Additionally if the experimental energy resolution were to exceed the neutrino mass scale, then an exceptionally good control of systematic uncertainties would also be required. 
For nonrelativistic neutrinos, the Majorana wind signal is suppressed relative to the Dirac case by the cancellation of the leading helicity-odd angular-correlation term, leading parametrically to an exposure penalty of order $(m_\nu/T_\nu)^2$.
\newline \newline 
\end{minipage}
}

\maketitle{}
\thispagestyle{empty}
\addtocounter{page}{-1}
\clearpage
\noindent
\hrule
\tableofcontents
\noindent
\hrulefill

\renewcommand{\thefootnote}{\arabic{footnote}}
\setcounter{footnote}{0}

\newpage

\section{Introduction}
\label{sec:intro}

The $\Lambda\mathrm{CDM}$ concordance model of cosmology predicts that neutrinos were emitted from the primordial plasma in the early universe and that a population of these neutrinos survive in the universe today as a cosmological relic known as the cosmic neutrino background (CNB)~\cite{Dolgov:2002wy,Lesgourgues:2014zoa,Akita:2022hlx}.  
Cosmological observables already furnish valuable information about the CNB.  
The cosmic microwave background (CMB) radiation's temperature and polarization anisotropies~\cite{Planck:2018vyg} bear the subtle imprint of the CNB neutrinos~\cite{Bashinsky:2003tk,Follin:2015hya}.  
These measurements reveal that the Universe contained $N_\mathrm{eff} \approx 3$ species of relativistic CNB neutrinos at the time of recombination when the Universe was approximately $300,\!000$ years old.  
On the other hand, galaxy surveys that probe the gravitational clustering of matter at late times have been thus far unable to reveal the CNB's influence.  
In fact recent measurements~\cite{DESI:2025zgx,Elbers:2025vlz} curiously seem to favor the absence of CNB neutrinos in the Universe today~\cite{Craig:2024tky,Green:2024xbb,Graham:2025dqn}.  
This apparent inconsistency of cosmological observables in the context of $\Lambda$CDM cosmology motivates a renewed attention to laboratory observables, which offer a complementary pathway for understanding the CNB \cite{Gerbino:2022nvz}.  
One promising strategy for a direct detection of the CNB neutrinos in laboratories on Earth \cite{Weinberg:1962zza} entails measuring the electron spectrum endpoint of a $\beta$-decaying isotope (such as tritium) at high precision ($\Delta E \lesssim 0.1 \; \mathrm{eV}$) and at large exposure ($\Xi \approx 100 \; \mathrm{g} \, \mathrm{yr}$) \cite{Cocco:2007za,Lazauskas:2007da,Blennow:2008fh,Li:2010sn,Faessler:2011qj,Long:2014zva,Akita:2020jbo,Akita:2021hqn,Bauer:2022lri}.  
Work is underway \cite{Betts:2013uya,PTOLEMY:2019hkd,PTOLEMY:2024boh} to address the technical challenges~\cite{Cheipesh:2021fmg,Nussinov:2021zrj,PTOLEMY:2022ldz,Tan:2022eke,Casale:2025vgd,Apponi_2026} presented by such a bold proposal.  
Despite the impediments, it cannot be understated that a direct detection of CNB neutrinos in the laboratory would open a new window on the early universe and physics beyond the Standard Model.   

Setting aside practical considerations (for the moment), it is interesting to consider the wealth of information that lies untapped in the CNB.  
Here we enumerate several CNB properties and questions: 
\begin{itemize}
	\item  \textbf{Local flux.}  What is the flux of relic neutrinos at the Earth?  How is the flux shared by the three neutrinos?  Does the flux match the $\Lambda$CDM prediction?  Is it larger, perhaps due to a late-time source of nonrelativistic neutrinos?  Is it smaller, perhaps due to neutrino decay or annihilation?  The authors of Ref.~\cite{KATRIN:2022kkv,Bondarenko:2023ukx,delCastillo:2025qnr} study large local neutrino over-densities.  
	\item  \textbf{Rest frame.}  In what frame of reference do the CNB neutrinos have zero average velocity?  Do all three mass eigenstates share a common rest frame?  Does the CNB rest frame coincide with the CMB rest frame?  What is the relative velocity of the CNB rest frame and the Earth rest frame?  In other words, what what is the velocity ${\bm v}_w$ of the CNB wind?  
	\item  \textbf{Energy/momentum spectra.}  How are the CNB neutrinos distributed over energy and over momentum?  Are these two distributions consistent with the usual energy-momentum relation $E = \sqrt{p^2 + m^2}$, which has only been probed for relativistic neutrinos, or do they differ?  Do these distributions match the thermal spectrum predicted by $\Lambda$CDM?  Are there spectral distortions?  The authors of Ref.~\cite{Chen:2015dka} studied the signatures of a nonthermal population of CNB relic neutrinos. The authors of Ref.~\cite{Huang:2021zzz} studied a model with a modified neutrino dispersion relation. 
    The authors of Refs.~\cite{Barenboim:2025vrc,Barenboim:2025vme} studied cosmological constraints on phenomenological parameterizations of CNB spectral distortions. 
	\item  \textbf{Polarization.}  Is the CNB polarized?  What are the relative fluxes of positive- and negative-helicity neutrinos?  Are the polarization fractions the same for all three masses?  The authors of Ref.~\cite{Baym:2020riw,Baym:2021ksj,Hernandez-Molinero:2022zoo} studied how polarization varies as neutrinos propagate through cosmic structures. 
	\item  \textbf{Temporal variation.}  How do the fluxes, spectra, and polarization vary with time?  The authors of Ref.~\cite{Safdi:2014rza,Zimmer:2025ohu} studied how solar gravitational focusing induces an annual modulation.  
	\item  \textbf{Anisotropies.}  How do the fluxes, spectra, and polarization vary across the sky? The authors of Refs.~\cite{Michney:2006mk,Hannestad:2009xu} studied how anisotropies should develop, and the authors of Ref.~\cite{Lisanti:2014pqa} studied how a polarized target could be used to access dipolar anisotropies. 
\end{itemize}
Questions abound regarding the properties of the CNB, and the answers to these questions have the potential to reveal a wealth of information about the physics of neutrinos and their role in cosmology.  

Now returning to the realm of practical considerations, the preceding list is roughly ordered in terms of increasing experimental difficulty.  
Although measuring the local flux of CNB neutrinos will be challenging, measuring their polarization anisotropies would be nigh impossible.  
Nevertheless, it is interesting to ask how we can ascribe a quantitative measure of difficulty to each CNB property, so as to better understand how far we are from reaching these goals.  
For instance, is it more accurate to say that measuring the anisotropies is $2$ times harder than measuring the flux, or $100$ times harder?  

As a first step, in this work we focus on the CNB wind.  
We seek to quantify the difficulty of detecting the CNB wind by measuring the angular distribution of recoiling electrons at the tritium beta decay endpoint.\footnote{
See also Refs.~\cite{Stodolsky:1974aq,Domcke:2017aqj,Shergold:2021evs}, in which the authors calculate how the CNB wind exerts a mechanical acceleration on a pendulum, such as ones found in gravitational wave interferometer experiments or torsion balances.  
}
If the CNB shares a common rest frame with the CMB, which is the $\Lambda$CDM prediction, then the relative speed of a laboratory on Earth and the CNB rest frame is typically $v_w \sim 10^{-3} \, c$, where the subscript $w$ stands for ``wind''.
As we discuss further in Sec.~\ref{sec:signal}, this leads to a dipolar modulation of the flux $\Phi_\nu(\hat{\bm n}) \sim \Phi_{\nu,0} + \delta\Phi_\nu \, \cos\theta$ where the relative amplitude of the dipole is $\delta\Phi_\nu / \Phi_{\nu,0} \sim v_w / v_\nu \sim 0.5 (v_w / 10^{-3} \, c) (m_\nu / 0.1 \; \mathrm{eV}/c^2) (T_\nu / 0.2 \; \mathrm{meV}/k_B)^{-1}$.  
The neutrino flux anisotropy can be large, particularly for heavier neutrinos that are slower in the CNB frame.  
However, we will see that the resultant anisotropy in the rate of recoiling electrons is much smaller: $\delta \Gamma_{\rm CNB} / \Gamma_{{\rm CNB},0} \sim $ $10^{-3}$ for nonrelativistic Dirac neutrinos with $m_\nu \sim 0.1 \, {\rm eV}$, and even smaller for Majorana neutrinos. This suppression arises from the small outgoing-electron velocity, $v_e/c\ll 1$, together with a partial cancellation in the nuclear form-factor combination that controls the electron–neutrino angular correlation.  
To assess the feasibility of detecting this angular variation, we consider an experimental setup with a mediocre energy resolution, such that the signal is background dominated, and with an enormous exposure, such that the signal-to-background ratio becomes observably large.  
We take the exposure $\Xi$ as our measure of the difficulty of detecting the CNB wind.  
The authors of Ref.~\cite{Betts:2013uya} have estimated that an exposure on the order of $\Xi \approx 100 \; \mathrm{g} \, \mathrm{yr}$ would be required to detect the CNB flux, and we seek to know how much additional exposure would be required to detect the CNB wind.

The remainder of this paper is organized as follows. 
In Sec.~\ref{sec:signal} we present our calculation of the dipolar anisotropy in the signal rate arising from the CNB wind, and we estimate that a large exposure would be required to reach $O(1)$ expected events, even in the absence of noise.  
In Sec.~\ref{sec:observables} we discuss irreducible backgrounds (beta decay) and experimental limitations (energy resolution), and we propose a strategy for extracting the CNB wind signal using large exposure.  
In Sec~\ref{sec:statistics} we present our statistical framework, including the treatment of nuisance parameters and the construction of Asimov sensitivities.
In Sec.~\ref{sec:results} we present our main results, including a figure showing the required exposure for detection of the CNB wind under different assumptions about the neutrino mass scale, the neutrino Dirac/Majorana nature, and the experimental energy resolution. 
We conclude in Sec.~\ref{sec:conclusions} with a short summary and outlook.

\section{Dipolar anisotropy from CNB wind}
\label{sec:signal}

In this section we present our calculation of the dipolar anisotropy that arises from the CNB wind.  

\subsection{CNB momentum distribution}
\label{sec:distribution function}

The $\Lambda$CDM cosmology predicts that neutrinos maintained thermal equilibrium with the primordial plasma via the Standard Model weak interactions.  
As the plasma cooled below a temperature of $T_{\rm dec} \sim {\rm MeV}$, neutrino annihilation and production channels fell out of thermal equilibrium.  
The residual neutrinos subsequently free streamed, losing momentum due to cosmological redshift in the expanding Universe.  
These neutrinos survive in the Universe today, and they are called the cosmic neutrino background \cite{Dolgov:2002wy,Mangano:2005cc,Lesgourgues:2012uu}.  

To derive predictions for relic neutrino direct detection, one must know the momentum distribution of the CNB neutrinos at Earth today.  
This distribution is informed by three things \cite{Hagmann:1999kf}:  (1) the intrinsic thermal spectrum of relic neutrinos, (2) the relative velocity of the Earth and the CNB rest frame, and (3) gravitational effects.\footnote{
Examples of gravitational effects include gravitational clustering \cite{Ringwald:2004np,Honma:2012zd,deSalas:2017wtt,Zhang:2017ljh,Mertsch:2019qjv,Holm:2023rml,Zimmer:2023jbb,Nascimento:2023ezc,Elbers:2023mdr,Holm:2024zpr,Zimmer:2024max} and gravitational focusing by the Sun \cite{Safdi:2014rza,Zimmer:2025ohu}.  
The change in the phase-space distribution due to gravitational clustering is $O(10\,\%)$ for $m_\nu\sim 0.1\,{\rm eV}$ assuming an NFW dark matter density profile.  
However, it strongly depends on the dark matter halo models adopted \cite{deSalas:2017wtt,Zhang:2017ljh,Mertsch:2019qjv}.
As our goal is to provide an order-of-magnitude estimate of the detectability, we neglect gravitational effects in our analysis.
}
The frame in which the average CNB neutrino momentum is equal to zero is known as the CNB rest frame.  
If the CNB rest frame coincides with the CMB rest frame, as $\Lambda$CDM cosmology predicts, then the relative velocity ${\bm v}_w$ of the Earth and the CNB rest frame has $v_w = |{\bm v}_w| \approx 370 \, \mathrm{km} \, {\rm s}^{-1} \approx 1.2 \times 10^{-3} \, c$ \cite{Planck:2018lkk,Saha:2021bay}.  
In the Earth's frame, the average number density of neutrinos with momentum between ${\bm p}_\nu$ and ${\bm p}_\nu + {\rm d} {\bm p}_\nu$ is given by ${\rm d} n_\nu = f_\nu({\bm p}_\nu) \, {\rm d}^3 {\bm p}_\nu / (2\pi)^3$.  
The phase-space occupation number per helicity degree of freedom is approximated as (also see Appendix~\ref{sec:derivation of distribution function})
\begin{align}
    f_\nu({\bm p}_\nu) 
    = f_0(|{\bm p}_\nu - E_\nu {\bm v}_w|)  
    + O\bigl( m_\nu^{2} / T_{\rm dec}^{2} \bigr) 
    \label{eq:fCNB}
\end{align}
where $E_\nu = \sqrt{|{\bm p}_\nu|^{2} + m_\nu^{2}}$ for neutrinos of mass $m_\nu$ and $v_w/c\ll 1$. 
Since $m_\nu / T_{\rm dec} \lesssim 10^{-10}$ for $m_\nu \lesssim 0.1 \, \mathrm{eV}$, the subleading term is completely negligible. 
In this expression, 
\begin{align}
    f_0(p_\nu)\equiv \frac{1}{\exp(p_\nu/T_{\nu})+1}\,,\label{eq:f0}
\end{align}
is the Fermi-Dirac phase space distribution function at temperature $T_{\nu} \simeq (4/11)^{1/3} \, T_{\rm CMB} \approx 0.168 \, {\rm meV}$ where $T_{\rm CMB} = 2.72548 \pm 0.00057 \, {\rm K}$ \cite{Fixsen_2009}.  

\begin{figure}[t]
\centering
\includegraphics[width=0.70\linewidth]{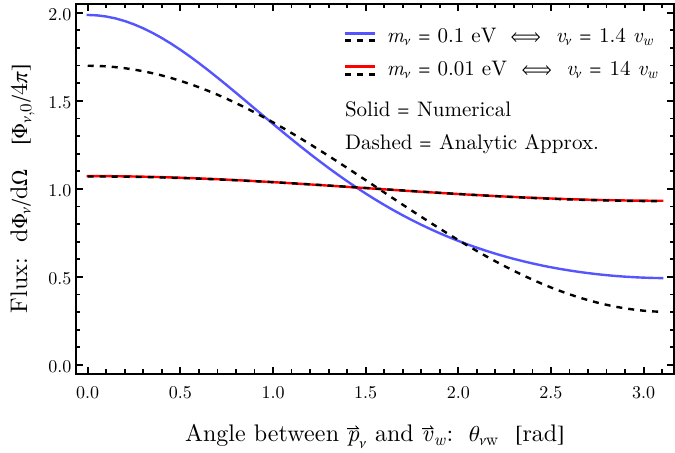}
\caption{\label{fig:flux}
Differential flux of incident neutrinos.  Solid curves are calculated by evaluating the integral in \Eref{eq:Phi_nu} numerically, and dashed curves are calculated by evaluating the expressions in \Eref{eq:Phi_nu_dipole_approx}.  
}
\end{figure}

The differential flux of CNB neutrinos is calculated as ${\rm d} \Phi_\nu = f_\nu({\bm p}_\nu) \, {\rm d}^3{\bm p}_\nu/(2\pi)^3 \times |{\bm p}_\nu| / E_\nu$, implying 
\begin{align}\label{eq:Phi_nu}
    \frac{{\rm d} \Phi_\nu}{{\rm d}\Omega_\nu} = \bar{v}_{\nu}T_\nu^3 \int_0^\infty \! \! \frac{x^3 {\rm d}x}{(2\pi)^3}\frac{m_\nu}{E_\nu} \, \biggl[ \mathrm{exp}\biggl( \sqrt{x^2 -  2 x\tfrac{v_w}{\bar{v}_\nu}\tfrac{E_\nu}{m_\nu} \cos\theta_{\nu w} + \tfrac{v_w^2}{\bar{v}_\nu^2}\tfrac{E_{\nu}^{2}}{m_\nu^{2}}} \biggr) + 1 \biggr]^{-1} 
    \;,
\end{align}
where the solid angle $\Omega_\nu$ indicates the direction of ${\bm p}_\nu$, and where $\theta_{\nu w}$ is the angle between ${\bm p}_\nu$ and ${\bm v}_w$.  
In this expression, we've defined $\bar{v}_\nu \equiv T_\nu / m_\nu$, and note that $\bar{v}_\nu \approx (1.68 \times 10^{-3} \, c) \, (m_\nu / 0.1 \, \mathrm{eV})^{-1} \approx (1.4 \, v_w) \, (m_\nu / 0.1 \, \mathrm{eV})^{-1}$.  
Note that $\langle |{\bm p}_\nu|/m_\nu \rangle = 7 \pi^4 \bar{v}_\nu / 180 \zeta(3) \approx 3.15 \, \bar{v}_\nu$ and $\sqrt{ \langle |{\bm p}_\nu|^2 / m_\nu^2 \rangle} = \sqrt{15 \zeta(5) / \zeta(3)} \, \bar{v}_\nu \approx 3.60 \, \bar{v}_\nu$ in the CNB rest frame. 
For $T_{\nu} \ll m_\nu \ll 0.1 \, \mathrm{eV}$ such that $v_w \ll \bar{v}_\nu \ll 1$, we can approximate 
\begin{subequations}\label{eq:Phi_nu_dipole_approx}
\begin{align}
    \frac{{\rm d} \Phi_\nu}{{\rm d}\Omega} \approx \frac{\Phi_{\nu,0}}{4\pi} + \frac{\delta\Phi_\nu}{4\pi} \, \cos\theta_{\nu w} + O(v_w^2 / \bar{v}_\nu) 
    \label{eq:dipole approximation of flux}
    \;,
\end{align}
where
\begin{align}
    \Phi_{\nu,0} & = \frac{7 \pi^2}{240} \, \bar{v}_\nu T_\nu^3 \approx \bigl( 8.95 \times 10^9 \, \mathrm{cm}^{-2} \, \mathrm{sec}^{-1} \bigr) \biggl( \frac{m_\nu}{0.1 \, \mathrm{eV}} \biggr)^{-1} \biggl( \frac{T_\nu}{0.168 \, \mathrm{meV}} \biggr)^{4} 
    \label{eq:analytic phi_nu0}
    \\ 
    \delta\Phi_{\nu} & = \frac{9 \zeta(3)}{4\pi^2} \, v_w T_\nu^3 \approx \bigl( 5.07 \times 10^9 \, \mathrm{cm}^{-2} \, \mathrm{sec}^{-1} \bigr) \biggl( \frac{v_w}{10^{-3}} \biggr)^{} \biggl( \frac{T_\nu}{0.168 \, \mathrm{meV}} \biggr)^{3} 
    \;.
    \label{eq:analytic delta phi_nu}
\end{align}
\end{subequations}
These two terms represent the isotropic component of the CNB flux ($\Phi_{\nu,0}$) and the dipolar modulation of the CNB flux ($\delta \Phi_\nu$).  
Fig.~\ref{fig:flux} shows the differential flux of incident neutrinos ${\rm d}\Phi_\nu / {\rm d}\Omega_\nu$ as a function of the angle $\theta_{\nu w}$ between the neutrino momentum ${\bm p}_\nu$ and the wind velocity ${\bm v}_w$.  
For $m_\nu = 0.1 \, {\rm eV}$ there is an $O(1)$ angular modulation, which is reasonably well approximated as a dipole.  
For $m_\nu = 0.01 \, {\rm eV}$ there is a tiny angular modulation (i.e., nearly isotropic), which is very well approximated as a dipole. 
For $m_\nu \lesssim T_\nu \approx 10^{-4} \, {\rm eV}$, the dipolar approximation is still reliable, but the analytical expression in \eqref{eq:analytic phi_nu0} is changed.
Next we study how the dipolar anisotropy in the incident neutrino flux translates into a dipolar anisotropy in the recoiling electron rate.

\subsection{Neutrino capture rate}
\label{sec:neutrino capture rate}

A leading proposal for the direct detection of CNB neutrinos involves their capture on a beta-decaying isotope \cite{Weinberg:1962zza,Cocco:2007za}.  
We consider the reaction 
\begin{align}\label{eq:reaction}
    \nu_i + \Tritium \to \He + e^{-}\,,
\end{align}
in which a neutrino of mass $m_{\nu_i}$ is incident on a tritium nucleus, which converts into a helium-3 nucleus accompanied by the emission of an electron.  
Since a nonrelativistic CNB neutrino carries an energy of approximately $m_{\nu_i}$, the recoiling electron is emitted with a kinetic energy that is approximately $2m_{\nu_i}$ above the beta decay endpoint ($Q_\beta \approx 18.6 \, \mathrm{keV}$).  
This is the characteristic signal of CNB capture. 

We focus on a single mass eigenstate and suppress the index $i$.
The differential rate for CNB neutrino capture on a single $\Tritium$ nucleus in its rest frame is calculated as 
\begin{align}
    \begin{split}
        \D \Gamma_{\rm CNB}^{X}\equiv \D\Pi_{\nu}\D\Pi_{e}\D\Pi_{\He}\frac{f_\nu({\bm p}_\nu)(|\mathcal{M}|^{2})^{X}}{2^{4}m_{\Tritium}E_{\He}E_{e}E_{\nu}} \, F(Z,E_{e}) \, (2\pi)^{4}\delta^{(4)}(p_\nu+p_{\Tritium}-p_{\He}-p_e)\,.
    \end{split}
    \label{eq:dGamma}
\end{align}
In this expression, ${\bm p}_\nu$ is the 3-momentum of the incident neutrino, $p_\nu^\mu = (E_\nu, {\bm p}_\nu)$ is the associated 4-momentum with energy $E_\nu = \sqrt{|{\bm p}_\nu|^2 + m_\nu^2}$, $m_{\Tritium}\simeq 2808.92\,{\rm MeV}$ is the mass of the incident tritium nucleus (at rest), ${\bm p}_e$ is the 3-momentum of the recoiling electron, $p_e^\mu = (E_e, {\bm p}_e)$ is the associated 4-momentum with energy $E_e = \sqrt{|{\bm p}_e|^2 + m_e^2}$ and mass $m_e \approx 0.511 \, \mathrm{MeV}$, ${\bm p}_{\He}$ is the 3-momentum of the recoiling helium-3 nucleus, $p_{\He}^\mu = (E_{\He}, {\bm p}_{\He})$ is the associated 4-momentum with energy $E_{\He} = \sqrt{|{\bm p}_{\He}|^2 + m_{\He}^2}$ and mass $m_{\He}\simeq 2808.39\,{\rm MeV}$.  
The Dirac delta function enforces energy and 3-momentum conservation.  
The neutrino momentum distribution function $f_\nu({\bm p}_\nu)$ is given by \Eref{eq:fCNB}. 
The Fermi function \cite{Fukugita:2003en},  
\begin{align}
    F(Z,E_{e}) = \frac{2\pi\eta}{1-e^{-2\pi\eta}} 
    \qquad \text{where} \qquad 
    \eta = Z \alpha E_e / |{\bm p}_e| 
    \;,
\end{align}
accounts for the enhanced scattering probability due to the Coulomb attraction between the emitted electron and the daughter nucleus.  
Here $Z=2$ is the atomic number of the daughter $\He$ nucleus, and $\alpha\simeq 1/137.036$ is the electromagnetic fine structure constant. 

The spin-dependent squared matrix element is given by \cite{Long:2014zva}
\begin{align}
    |\mathcal{M}|^{2}(s_{\nu}) = 8 G_{F}^{2} |U_{ei}|^{2} |V_{ud}|^{2} m_{\Tritium} E_{\He} E_{\nu} E_{e} \, \bigl[ C_{A} \, A(s_{\nu}) + C_{B} \, B(s_{\nu})v_e\cos\theta_{e\nu}\bigr] 
    \,,
    \label{eq:matrix element}
\end{align}
where $G_F \approx 1.166 \times 10^{-5} \, {\rm GeV^{-2}}$ is Fermi's constant, $|U_{ei}| \approx 0.82$ is the $(e,i)$ component of the PMNS matrix, and $|V_{ud}| \approx 0.97$ is the $(u,d)$ component of the CKM matrix.  

The nuclear matrix elements enter via~\cite{ParticleDataGroup:2024cfk,Schiavilla:1998je} 
\begin{align}
    C_{A}=\langle f_F\rangle^{2}+(g_A/g_V)^{2}\langle g_{GT}\rangle^{2}\simeq 5.49\,,\quad 
    C_{B}=\langle f_F\rangle^{2}-\frac{1}{3}(g_A/g_V)^{2}\langle g_{GT}\rangle^{2}\simeq -0.50\,,\label{eq:FA and FB}
\end{align}
where $\langle f_F\rangle\simeq 0.9987$, $g_A\simeq 1.2695$, $g_{V}\simeq 1$, and $\langle g_{GT}\rangle^{2}\simeq 2.788$.
In the last term $v_e = |{\bm p}_e| / E_e$ is the speed of the recoiling electron, 
and $\theta_{e\nu}$ is the angle between the directions of the incoming neutrino and the outgoing electron momenta.

The squared matrix element depends on the helicity of the incident neutrino ($s_{\nu} = \pm 1/2$) via \cite{Long:2014zva}
\begin{align}
    A(s_{\nu}) & \equiv 1 - 2 s_{\nu} v_{\nu} 
    = \begin{cases} 1 - v_{\nu} & , \quad s_{\nu} = +1/2 \\ 1 + v_{\nu} & , \quad s_{\nu} = -1/2 \end{cases} \\ 
    B(s_{\nu}) & \equiv - 2 s_{\nu} + v_{\nu} 
    = \begin{cases} -1 + v_{\nu} & , \quad s_{\nu} = +1/2 \\ 1 + v_{\nu} & , \quad s_{\nu} = -1/2 \end{cases} 
    \;,
    \label{eq:spin-dependent factor AB}
\end{align}
where $v_{\nu} = |{\bm p}_{\nu}| / E_{\nu}$ is the speed of the incident neutrino.  
The way that these factors enter observables depends on whether the light neutrinos are Dirac or Majorana. 
In the early universe, neutrinos/anti-neutrinos maintained thermal equilibrium via the Standard Model weak interactions, which are chiral.  
This led to a thermal population of negative helicity neutrinos and positive helicity anti-neutrinos.  
After neutrino decoupling, their helicity was approximately conserved during the cosmological expansion \cite{Baym:2020riw,Baym:2021ksj,Hernandez-Molinero:2022zoo}, even after becoming nonrelativistic.  
Therefore, if the light neutrinos/antineutrinos are the quantum excitations of a Dirac spinor field, then the CNB consists of only negative helicity neutrinos and positive helicity antineutrinos, but the latter are undetectable via neutrino capture on tritium.  
Conversely, if the light neutrinos are the quantum excitations of a Majorana spinor field, then the CNB consists of both negative and positive helicity neutrinos, which are all available for capture on tritium.  
To account for these two cases we define 
\begin{align}
    (|\mathcal{M}|^2)^X = 
    \begin{cases}
    (|\mathcal{M}|^2)^D = |\mathcal{M}|^2(s_{\nu} = -1/2) & , \quad \text{Dirac neutrinos} \\ 
    (|\mathcal{M}|^2)^M = |\mathcal{M}|^2(s_{\nu} = -1/2) + |\mathcal{M}|^2(s_{\nu} = +1/2) & , \quad \text{Majorana neutrinos} 
    \end{cases} 
    \;.
\end{align}
It is also useful to define 
\begin{subequations}
\begin{align}
    A^X & = \begin{cases}
    A^{D} = A(-1/2) = 1+v_{\nu} & , \quad \text{Dirac} \\ 
    A^{M} = A(+1/2) + A(-1/2) = 2 & , \quad \text{Majorana} 
    \end{cases} \\ 
    B^X & = \begin{cases}
    B^{D} = B(-1/2) = 1+v_{\nu} & , \quad \text{Dirac} \\ 
    B^{M} = B(+1/2) + B(-1/2) = 2 v_{\nu} & , \quad \text{Majorana} 
    \end{cases} 
    \;,
\end{align}
\end{subequations}
which enter the expressions for $(|\mathcal{M}|^2)^D$ and $(|\mathcal{M}|^2)^M$.  

For CNB capture on tritium \eqref{eq:reaction}, the recoiling helium-3 nucleus is unobserved, and one must infer the presence of an incident CNB neutrino using only the recoiling electron.  
Thus we are mainly interested in the differential event rate $\D\Gamma_\mathrm{CNB}^X / \D E_e \D\Omega_e$, which gives the distribution over the electrons' energies and orientations. 
Next we briefly review the nearly-monochromatic energy spectrum, and afterward we focus on the angular distribution for the remainder of this section. 

To understand the electron's spectrum, it suffices to consider the kinematics of the 2-to-2 neutrino capture reaction \eqref{eq:reaction} in the frame where the incident tritium nucleus is at rest.  
Energy and momentum conservation imply that the recoiling electron is emitted with an energy $E_e \approx E_\ast$ and a momentum $p_e \approx p_\ast$, which are given by \cite{Long:2014zva} 
\begin{align}
    E_* \equiv \frac{(m_{\Tritium}+E_\nu)^{2}+m_e^{2}-m_{\He}^2}{2(m_{\Tritium}+E_\nu)} 
    \qquad \text{and} \qquad 
    p_*=\sqrt{E_*^2 - m_e^{2}}\,.
    \label{eq:starred Ee}
\end{align}
These relations assume that no energy is lost to vibrational or rotational excitations of the daughter helium-3 nucleus, 
and they approximate $p_\nu \ll m_{\He}$.  
The electron energy $E_{*}$ depends on the incident neutrino momentum $p_\nu$ through $E_\nu = (p_\nu^2 + m_\nu^2)^{1/2}$. Since $p_\nu$ is drawn from a thermal distribution with width $\sim T_\nu \approx 0.168 \, \mathrm{meV}$, the electron spectrum inherits a thermal broadening.  However, this intrinsic width is generally much smaller than the experimental energy resolution, say $50 \, \mathrm{meV}$, which we discuss further in the next section.  Therefore, it suffices to approximate the electron spectrum as a monochromatic line, 
\begin{align}
    \frac{\D ^{2}\Gamma_{\rm CNB}^{X}}{\D E_{e}\D \Omega_e}
    \simeq
    \frac{\D \Gamma_{\rm CNB}^{X}}{\D \Omega_e} \, 
    \delta(E_e-E_\ast)\;,
\end{align}
where $E_\ast \approx m_e + K_{\rm end}^0 + m_\nu$ and $K_{\rm end}^0 \approx 18.5988 \, \mathrm{keV}$.

Finally we turn our attention to the electron's angular distribution, given by the differential event rate $\D\Gamma_{\rm CNB}^X / \D\Omega_e$.  
In order to derive a simple analytical expression, we employ three approximations.  

For the first approximation, we neglect recoil-suppressed corrections of order $E_e/E_{^3{\rm He}}$.  Under this approximation, the angular dependence of the phase-space integral simplifies.  In particular, as shown explicitly in Appendix~\ref{sec:detection rate}, the angular dependence of the Jacobian can be neglected, and the only remaining dependence on the incoming-neutrino direction comes from the squared matrix element. Since \Eref{eq:matrix element} contains only an isotropic term and a term linear in $\cos\theta_{e\nu}$, the angular integration of $|\mathcal{M}|^2 f_\nu$ selects only the monopole and dipole moments of the boosted CNB distribution by the orthogonality of Legendre polynomials.  Therefore, up to corrections of order $E_e/E_{^3{\rm He}}\sim 10^{-4}$, the electron angular distribution can be written as
\begin{align}
    \frac{\D\Gamma_{\rm CNB}^{X}}{\D\Omega_{e}} 
    \simeq 
    \frac{\Gamma_{{\rm CNB},0}^X}{4\pi} + \frac{\delta\Gamma_{\rm CNB}^X}{4\pi} \, \cos\theta_{ew} \;,
    \label{eq:dGamma_CNB}
\end{align}
where $\theta_{ew}$ is the angle between $\bm p_e$ and $\bm v_w$.
Here $\Gamma_{{\rm CNB},0}^X$ is the isotropic component of the rate and $\delta\Gamma_{\rm CNB}^X$ is the amplitude of the dipole component.  

For the second approximation, we use $p_\nu\ll p_e$. This approximation implies that the kinematically allowed range of the daughter momentum is restricted to a narrow interval around $p_{^3{\rm He}}\in[p_\ast-p_\nu,p_\ast+p_\nu]$.
Over this narrow interval, factors such as $E_e$ $p_e$, and $F(Z,E_e)$ vary only weakly.  We therefore evaluate them at $E_e=E_\ast$ and $p_e=p_\ast$ when performing the $p_{^3{\rm He}}$-integral.  This approximation is accurate up to corrections of order $p_\nu/p_e\sim T_\nu/p_e\sim 10^{-9}$, which are much smaller than the recoil correction discussed above. 
With these approximations, the isotropic and dipole components of \Eref{eq:dGamma_CNB} are written as 
\begin{align}
\begin{split}
    \Gamma_{{\rm CNB},0}^{X} & =  4\pi \biggl( \frac{G_{F}^{2}|U_{ei}|^{2}|V_{ud}|^{2}}{(2\pi)^{4}} \biggr) C_A \mathcal{I}_{A}^{X} \label{eq:Gamma_CNB} \\ 
    \delta\Gamma_{\rm CNB}^{X} & = 4\pi \biggl( \frac{G_{F}^{2}|U_{ei}|^{2}|V_{ud}|^{2}}{(2\pi)^{4}} \biggr) \tfrac{1}{3} C_B \mathcal{I}_{B}^{X} 
    \;,
    \end{split}
\end{align}
where the trailing factors are defined to be the integrals 
\begin{align}
\begin{split}
    \mathcal{I}_A^X & \simeq \int_0^{\infty} \D p_\nu \, p_\nu^{2} \, f_0(p_\nu) \, A^{X}(v_\nu) \, E_* p_* \, F(Z,E_*) 
    \;,\\ 
    \mathcal{I}_B^X & \simeq \int_{0}^{\infty} \D p_\nu \, p_\nu^{2} \, f_1(p_\nu) \, B^{X}(v_\nu) \, p_*^2 \, F(Z,E_*) 
    \;.
    \label{eq:A,B for M,D}
    \end{split}
\end{align}
Here $f_0(p_\nu)=1/(\exp(p_\nu/T_\nu)+1)$ and $f_1(p_\nu)=\tfrac{3}{2}\int_{-1}^{1}\D\cos\theta f_\nu(\bm{p}_\nu)\cos\theta$ are the monopole and dipole moments of the boosted CNB distribution.

\begin{figure}[t]
\centering
\includegraphics[width=0.75\linewidth]{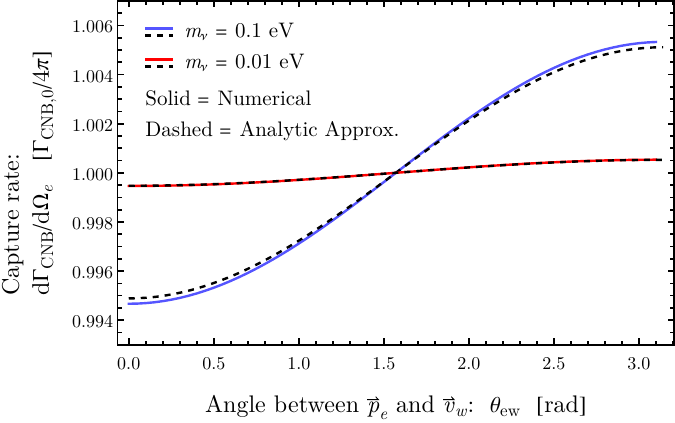}
\caption{\label{fig:rate}
Differential capture rate for Dirac neutrinos as a function of the angle between the recoiling-electron momentum ${\bm p}_e$ and the CNB wind velocity ${\bm v}_w$. We use the same benchmark parameters as in Fig.~\ref{fig:flux}. 
The solid curves are obtained by calculating the integrals in \Eref{eq:dGamma} using numerical methods.  
The dashed curves are obtained by evaluating the analytic approximation in \Erefs{eq:dGamma_CNB}–\eqref{eq:small boosted f1}.
}
\end{figure}

For the third approximation, to obtain a simple analytic expression for the integrals, we use the small-boost approximation $v_w \ll \bar{v}_\nu = T_\nu / m_\nu$. 
In this limit,
\begin{align}
    f_{1}(p_\nu)
\simeq
f_0(p_\nu)\left[1-f_0(p_\nu)\right] \frac{E_{\nu}}{m_\nu}\frac{v_w}{ \bar{v}_\nu}\,.
\label{eq:small boosted f1}
\end{align}
Together with the first two approximations, this gives the semi-analytic expression in \Eref{eq:A,B for M,D}. This is the expression used for the dashed curves in Fig.~\ref{fig:rate}; in evaluating it we keep the $p_\nu$-dependence of the kinematic factors such as $E_\ast(p_\nu)$ and $p_\ast(p_\nu)$.
Although the boosted distribution itself can contain higher multipoles when $v_w\not\ll\bar v_\nu$, these higher multipoles do not contribute to the capture rate once the first approximation above is made. Thus the relevant effect of the small-boost approximation is the analytic estimate of the dipole moment $f_{1}$, rather than the truncation of higher multipoles.

To assess the validity of these approximations, Fig.~\ref{fig:rate} shows a comparison of our analytical expressions and the outcome of direct numerical integration. 
We show only the Dirac case in Fig.~\ref{fig:rate}, since the Majorana dipole component is helicity-suppressed, as discussed below.
For the numerical calculation, we start with the definition of $\D\Gamma_{\rm CNB}^X$ in \Eref{eq:dGamma}.  The four Dirac delta functions are used to remove the integrals over $|{\bm p}_e|$ and ${\bm p}_{\He}$.  Since we are interested in $\D\Gamma_{\rm CNB}^X / \D\Omega_e$, we hold fixed the orientation of ${\bm p}_e$.  The remaining three integrals over ${\bm p}_\nu$ are evaluated numerically.
By inspecting the figure, we see that the analytical approximation and the direct numerical integration differ by no more than $\sim 0.02\%$ for the range of angles and for the set of parameters shown.  
This difference may be attributed to the recoil-suppressed correction of order $E_e/E_{^3{\rm He}}\sim 10^{-4}$, rather than to the finite width of the $p_{^3{\rm He}}$ integration region, whose effect is only of order $p_\nu/p_e\sim10^{-9}$.
The excellent agreement between analytical and numerical calculations gives us confidence to adopt the dipolar form of the differential event rate \eqref{eq:dGamma_CNB} in the remainder of this work. 

Before moving on, it is useful to further simplify \Eref{eq:A,B for M,D} in order to make the parametric dependence transparent.  This additional simplification is not used for the dashed curves in Fig.\ref{fig:rate}.  If the relic neutrinos are nonrelativistic, $T_\nu\ll m_\nu$, then $E_\nu\simeq m_\nu$ over the thermal distribution.  Consequently, the kinematic factors $E_\ast$, $p_\ast$, and $F(Z,E_\ast)$ may be treated as independent of $p_\nu$, while the helicity factors $A^X(v_\nu)$ and $B^X(v_\nu)$ may be evaluated at a representative velocity $\bar v_\nu=T_\nu/m_\nu$. The remaining radial integrals can then be performed analytically, giving
\begin{align}
\begin{split}
    \mathcal{I}_A^X & 
    \simeq   \frac{3}{2} \zeta(3) \, A^{X}(\bar{v}_\nu) \, T_\nu^3 \, E_\ast p_\ast \, F(Z,E_\ast)
    \;,\\ 
    \mathcal{I}_B^X & 
    \simeq  \frac{\pi^2}{6} B^{X}(\bar{v}_{\nu}) \, m_\nu T_\nu^2 \, p_*^2 \, F(Z,E_\ast) v_w
    \;,
    \label{eq:A,B for M,D, small boost}
    \end{split}
\end{align}
This nonrelativistic closed form will be used only to understand the scaling of the dipole-to-monopole ratio below.

%

It is illuminating to compare the dipole component of the rate with the isotropic component.  
Subject to the approximations listed above, this ratio is parametrically equal to 
\begin{align}
    \frac{|\delta\Gamma_{\rm CNB}^X|}{\Gamma_{{\rm CNB},0}^X} 
    \simeq  \frac{|C_B|}{3C_A} \frac{B^{X}}{A^{X}} \frac{\pi^2}{9 \zeta(3)} \,\frac{v_w}{\bar{v}_{\nu}} v_\ast 
    \;,
    \label{eq:RBA}
\end{align}
where $v_\ast = p_\ast / E_\ast$ and $\bar{v}_\nu = T_\nu / m_\nu$.  
A rough numerical estimate gives 
\begin{align}
    \frac{|\delta\Gamma_{\rm CNB}^X|}{\Gamma_{{\rm CNB},0}^X} 
    \sim 10^{-6} 
    \biggl( \frac{|C_B|/3C_A}{10^{-2}} \biggr) 
    \biggl( \frac{v_\ast}{10^{-1}} \biggr) 
    \biggl( \frac{v_w}{10^{-3}} \biggr) 
    \begin{cases}
    10^3 \, \bigl( \tfrac{m_\nu}{0.1 \, \mathrm{eV}} \bigr)^{} & , \quad \text{Dirac} \\ 
    1 & , \quad \text{Majorana} 
    \end{cases}
    \label{eq:RBA scaling}
\end{align}
A couple of remarks are in order.  

First, it is interesting to compare the rates for Dirac and Majorana neutrinos.  
Whereas the isotropic capture rates only differed by a factor of $A^D / A^M \approx 1/2$, the dipolar modulation differs by a factor of $B^D / B^M \approx 1 / \bar{v}_\nu \sim m_\nu / T_\nu$.  
For nonrelativistic neutrinos, this ratio may be as large as $\sim 10^{3}$, implying that Dirac neutrinos would exhibit a much larger dipolar modulation.  
The modulation is weaker for Majorana neutrinos, because there is a cancellation in the dipolar term when positive and negative helicities are summed.\footnote{
Recall $B^M = B(+1/2) + B(-1/2) = (-1 + v_\nu) + (1 + v_\nu) = 2 v_\nu$ whereas $B^D = B(-1/2) = (1 + v_\nu) \approx 1$. The angular correlation of the emitted electron contains a $\bm p_e\cdot\bm S_\nu$ spin-correlation term. For a helicity eigenstate, this produces the leading $O(1)$ contribution to $B(s_\nu)$, with opposite signs for the two helicities. Therefore, for Majorana neutrinos, the leading helicity-odd terms cancel when the two helicities are summed, and only the helicity-even $O(v_\nu)$ term remains.}

Note that this Dirac/Majorana separation assumes that the helicity composition of the CNB is preserved after decoupling, so that the Dirac CNB contains only negative-helicity neutrinos available for capture. While this is expected to be a good approximation in the standard cosmological setting, small helicity rotations induced by gravitational inhomogeneities or magnetic-field effects have been discussed in Refs.~\cite{Baym:2020riw,Baym:2021ksj,Hernandez-Molinero:2022zoo,Liao:2023zem,Peng:2022skd}. If such effects partially depolarize the CNB, the contrast between the Dirac and Majorana dipole amplitudes would be correspondingly modified.

Second, this estimate reveals that the dipolar modulation is small compared to the isotropic capture rate.  
The capture rate $\Gamma_{\rm CNB}^X$ (irrespective of electron energy or orientation) is calculated by integrating the differential rate $\D \Gamma_{\rm CNB}^X / \D\Omega_e$ over all possible orientations for the recoiling electron. 
For a tritium target with mass $M_{\Tritium}$, the expected number of events during time $\tau$ is calculated as 
\begin{align}
    N_{\rm CNB}^X = \Xi \, \Gamma_{\rm CNB}^X / m_{\Tritium} 
    \;,
\end{align}
where $\Xi = M_{\Tritium} \tau$ is the exposure.  
When the relevant mass eigenstates are unresolved and can be treated with a common representative mass, the PMNS weights combine to $\sum_i |U_{ei}|^2=1$. 
For $\Xi = 100 \, {\rm g} \, {\rm yr}$, we estimate $N_{\CNB}^M \simeq 8$ and $N_{\CNB}^D \simeq 4$, which is consistent with Ref.~\cite{Long:2014zva}.
Since the dipolar modulation is at the level of $\delta\Gamma_{\rm CNB}^D / \Gamma_{\rm CNB,0}^D \sim 10^{-3}$ for Dirac neutrinos and $10^{-6}$ for Majorana neutrinos, then a commensurately larger exposure would be needed to observe the dipolar modulation, even under ideal circumstances (no backgrounds, no detector noise).  
In the following sections, we quantify exactly how much exposure would be needed while also accounting for background-dominated measurements.

\section{Background contamination and energy resolution}
\label{sec:observables}

In this section we discuss the dominant background for CNB capture on tritium.  
Namely, tritium beta decays create electrons below the endpoint, but given the experiment's finite energy resolution, some of these electrons are measured above the endpoint in the CNB signal region.  
We discuss two analysis strategies that could be used to infer the CNB signal through its flux or dipolar modulation. 

\subsection{Tritium $\beta$-decay background}
\label{sec:tritum beta-decay background}
The dominant background in a tritium CNB experiment arises from tritium $\beta$-decay,
\begin{align}
    \Tritium\to \He + e^{-} + \bar{\nu}_{e}\,.
\end{align}
The corresponding differential decay rate per target $\Tritium$ atom is \cite{Cocco:2007za,Masood:2007rc}
\footnote{In a realistic implementation, the beta-decay background need not be exactly isotropic. For example, in substrate-based concepts such as PTOLEMY, where tritium atoms are bound to a two-dimensional material \cite{Betts:2013uya,PTOLEMY:2024boh}, the target geometry and electron transport could in principle induce an anisotropic acceptance or emission pattern. Modeling such detector- and target-dependent effects is beyond the scope of this work. In the following, we therefore adopt the idealized assumption that the beta-decay background is isotropic in the laboratory frame and write $( \D^2\Gamma_\beta / \D E_e\D\Omega_e ) = (1/4\pi) (\D\Gamma_\beta / \D E_e)$.
}
\begin{align}
    \frac{\D \Gamma_\beta}{\D E_e}
  = \sum_{i=1}^3 |U_{ei}|^2\,
    \frac{\bar{\sigma}}{\pi^{2}}H(E_e,m_{\nu_i})\,.
\end{align}
Here
\begin{align}
    H(E_e,m_{\nu_i})=\frac{1-m_{e}^{2}/(E_{e}m_{\Tritium})}{(1-2E_{e}/m_{\Tritium} + m_{e}^{2}/m_{\Tritium}^{2})^{2}}\sqrt{y\left(y+\frac{2m_{\nu_i}m_{\He}}{m_{\Tritium}}\right)}\left(y+\frac{m_{\nu_i}}{m_{\Tritium}}(m_{\He} + m_{\nu_i})\right),
\end{align}
with $y=(m_{e}+K_{\rm end}^{(i)}) -E_{e}$. The normalization factor is
\begin{align}
    \bar{\sigma}=\frac{G_{F}^{2}}{2\pi}|V_{ud}|^{2}F(Z,E_{e})\frac{m_{\He}}{m_{\Tritium}}E_{e}p_{e}F_A\,.
\end{align}
The factor $F_A$ is defined in \eqref{eq:FA and FB}.
The endpoint kinetic energy for the $i$-th neutrino mass eigenstate is
\begin{align}
    K_{\rm end}^{(i)}=\frac{(m_{\Tritium}-m_{e})^{2}-(m_{\nu_i} + m_{\He})^{2}}{2m_{\Tritium}}\simeq K_{\rm end}^0-m_{\nu_i}\,,
\end{align}
where $K_{\rm end}^{0}$ denotes the endpoint kinetic energy in the massless-neutrino limit.
Integrating over the electron energy, the total tritium $\beta$-decay rate per target $\Tritium$ atom is
\begin{align}
    \Gamma_\beta
    =
    \int _{m_e}^{m_e+K_{\rm end}^{0}}
    \D E_{e}
    \frac{\D \Gamma_\beta}{\D E_e}\,.
\end{align}
For an exposure $\Xi$, the expected number of $\beta$-decay events is $N_{\beta}=\Xi\Gamma_\beta/m_{\Tritium}\simeq 10^{24}\left(\frac{\Xi}{100\,{\rm g\cdot yr}}\right)$.
\footnote{In the following numerical estimates we use this leading endpoint spectrum, while residual subleading corrections such as radiative $\beta$-decay $\Tritium\to\He+e^{-}+\bar{\nu}_{e}+\gamma$ to the endpoint background are included in the nuisance treatment discussed in Sec.~\ref{sec:smearing} and Sec.~\ref{sec:statistics_energy_only}.}

\subsection{Modeling finite energy resolution}
\label{sec:smearing}

Following Ref.~\cite{Long:2014zva}, we account for the finite energy resolution by Gaussian smearing with the width $\sigma_E$,
\begin{align}
    \frac{\D ^{2}\tilde{\Gamma}_Y}{\D E_{e}\D \Omega_e}(E_{e},\Omega_{e})\equiv \int _{m_e}^{\infty}\D E_{e}'G(E_{e},E_{e}')\frac{\D ^{2}{\Gamma}_Y}{\D E_{e}\D \Omega_e}(E_{e}',\Omega_{e})\,,
\end{align}
where $Y={\rm CNB}$ or $\beta$. 
The Gaussian kernel is
\begin{align}
    G(E_{e},E_{e}')=\frac{1}{\sqrt{2\pi\sigma_{E}^{2}}}\exp\left(-\frac{(E_{e}'-E_{e})^{2}}{2\sigma_{E}^{2}}\right)\,.
\end{align}
The corresponding full width at half maximum (FWHM) is
\begin{align}
    \Delta E_{\rm eff}=2\sqrt{2\ln 2}\,\sigma_E\,.
\end{align}
Using the monochromatic approximation for the CNB signal, we obtain
\begin{align}
    \frac{\D ^{2}\tilde{\Gamma}_{\rm CNB}^{X}}{\D E_{e}\D \Omega_e}\simeq \frac{\D \Gamma_{\rm CNB}^{X}}{\D \Omega_e}G(E_{e},E_{\rm pk})
    \simeq\left(\frac{\Gamma_{\rm CNB,0}^{X}}{4\pi}+\frac{\delta\Gamma_{\rm CNB}^{X}}{4\pi}\cos\theta_{ew}\right)G(E_{e},E_{\rm pk})\,.
\end{align}
In the following analysis, we consider a single energy bin centered at $E_{e}=E_{\rm pk}$ with a width $\Delta E_{\rm eff}$. The angular differential rate in this signal bin is given by
\begin{align}
    \frac{\D \tilde{\Gamma}_{\rm CNB}^{X}}{\D \Omega_{e}}=\int _{L}^{U}\D E_{e}\frac{\D ^{2}\tilde{\Gamma}_{\rm CNB}^{X}}{\D E_{e}\D \Omega_e}\simeq \frac{\D \Gamma_{\rm CNB}^{X}}{\D \Omega_e}\int _{L}^{U}\D E_{e}G(E_{e},E_{\rm pk})
    \simeq\left(\frac{\Gamma_{\rm CNB,0}^{X}}{4\pi}+\frac{\delta\Gamma_{\rm CNB}^{X}}{4\pi}\cos\theta_{ew}\right)I_{G}\,.
\end{align}
Here 
\begin{align}
    I_G\equiv \int _L^U \D E_{e} \,G(E_{e},E_{\rm pk})\,,
\end{align}
with $L\equiv E_{\rm pk}-\Delta E_{\rm eff}/2$ and $U\equiv E_{\rm pk}+\Delta E_{\rm eff}/2$.

The effective energy resolution in tritium-on-graphene configurations receives contributions from the detector response, the quantum motion of tritium bound to the substrate, and the many-body final-state distribution of the $\He$ nucleus and the surrounding electronic degrees of freedom \cite{Nussinov:2021zrj,Cheipesh:2021fmg,PTOLEMY:2022ldz,Tan:2022eke,Casale:2025vgd,Apponi_2026}.  
Schematically, one may write
\begin{align}
    \sigma_E^2 = \sigma_{\rm eff}^{2}(\Delta E_{\rm det},\Delta E_{\rm init},\Delta E_{\rm FSD})\,,
\end{align}
where $\Delta E_{\rm det}$, $\Delta E_{\rm init}$, and $\Delta E_{\rm FSD}$ denote characteristic energy widths associated with detector response, initial-state motion, and final-state distributions, respectively.
Rather than modeling these mechanisms in detail (including possible non-Gaussian tails and correlations), we capture their net effect with a single effective Gaussian smearing characterized by $\Delta E_{\rm eff}$.

The effective endpoint background model may also receive subleading theoretical corrections to the $\beta$-decay spectrum, such as radiative $\beta$-decay, $\Tritium\to\He+e^{-}+\bar{\nu}_{e}+\gamma$. These corrections can change the normalization and shape of the smeared endpoint tail. We treat their residual impact, together with the response and final-state modeling uncertainties, as a single endpoint-normalization nuisance parameter with a Gaussian prior in Sec.~\ref{sec:statistics_energy_only}.

\subsection{Endpoint-rate analysis strategy}
\label{sec:SB in energy_only}

For the endpoint-rate analysis, (hereafter denoted as ``(rate)'' in the equations), we consider the total event count integrated over the angle in the signal bin as the observable. In other words, we test the nonzero CNB capture rate, defined as
\begin{align}
    \Gamma_{\rm S}^{({\rm rate}),X}\equiv I_{G}\Gamma_{\rm CNB,0}^{X}\,,
\end{align}
with the background rate, given by
\begin{align}
    \Gamma_{\rm B}^{({\rm rate})}\equiv \int _L^U \D E_{e}\int \D \Omega_e\,\frac{\D ^{2}\tilde{\Gamma}_{\beta}}{\D E_{e}\D \Omega_e}\,.
\end{align}
In this test, the signal-to-noise ratio is given by
\begin{align}
    r_{\rm S/B}^{({\rm rate}),X}=\frac{\Gamma^{({\rm rate}),X}_{\rm S}}{\Gamma_{\rm B}^{({\rm rate})}}\,.
    \label{eq:r_rate_E}
\end{align}

As shown in Ref.~\cite{Long:2014zva}, one obtains $r_{\rm S/B}^{({\rm rate}),X}\ge 1$ for $\Delta E_{\rm eff}\lesssim 0.7m_\nu$. For small neutrino masses this condition is stringent. If the lightest neutrino is (almost) massless, oscillation data imply that the heaviest mass eigenstate has mass at most $m_\nu \sim 50\,{\rm meV}$\cite{Esteban:2024eli}, and $\Gamma_{\rm S}^{({\rm rate}),X}\gtrsim \Gamma_{\rm B}^{({\rm rate})}$ requires $\Delta E_{\rm eff} \lesssim  35\,{\rm meV}$. Even in the optimistic case of quasi-degenerate masses $m_{\nu_1}\sim m_{\nu_2}\sim m_{\nu_3}$, cosmological observations give $\sum_i m_{\nu_i}  \lesssim 0.3\,{\rm eV}$\cite{Planck:2018vyg}, so that $m_{\nu_i} \lesssim 0.1\,{\rm eV}$ and $\Delta E_{\rm eff} \lesssim  70\,{\rm meV}$. 
As already discussed, achieving such a small effective smearing width is challenged by fundamental initial- and final-state effects, not only by detector technology~\cite{Cheipesh:2021fmg,Nussinov:2021zrj,PTOLEMY:2022ldz,Tan:2022eke,Casale:2025vgd,Apponi_2026}. With a more realistic energy resolution, e.g.,~$\Delta E_{\rm eff}=m_\nu=0.1\,{\rm eV}$, we obtain $r_{\rm S/B}^{({\rm rate}),M}=6.2\times 10^{-4}$. 

The expected total count in the energy bin is
\begin{align}
    \lambda^{({\rm rate}),X}=S^{({\rm rate}),X}+B^{({\rm rate})}
    \,,\label{eq:lambda_E}
\end{align}
where $S^{({\rm rate}),X}\equiv \Xi\Gamma^{({\rm rate}),X}_{\rm S}/m_{\Tritium}$ and $B^{({\rm rate})}\equiv\Xi\Gamma^{({\rm rate})}_{\rm B}/m_{\Tritium}$ denote the expected signal and background counts for exposure $\Xi$, respectively.

\subsection{Endpoint-dipole analysis strategy}
\label{sec:SB in energy_angle}

For the endpoint-dipole analysis (hereafter denoted as ``(dipole)'' in the equations), we consider the same single endpoint energy bin as in the previous section and divide the range $u \equiv u_{ew}=\cos\theta_{ew}\in[-1,1]$ into $N_\Omega$ bins of width $\Delta u=2/N_\Omega$. 
In this analysis, we test the nonzero amplitude of the dipole contribution to the CNB capture rate, given by
\begin{align}
    \Gamma_{\rm S}^{({\rm dipole}),X}
    \equiv
    \frac{1}{2}I_{G}\delta\Gamma_{\rm CNB}
    =
    \frac{2\pi}{3}
    I_G \biggl( \frac{G_{F}^{2}|U_{ei}|^{2}|V_{ud}|^{2}}{(2\pi)^{4}} \biggr) C_{B}\mathcal{I}_{B}^{X}
    \label{eq:Gamma_aniso}
\end{align}
The dipole contribution in the $j$-th angular bin is then proportional to $\Gamma_{\rm aniso}^{(\Omega),X}u_j$, where $u_j$ denotes the bin-averaged value of $u=\cos\theta_{ew}=u_{ew}$ in the $j$-th angular bin.

Throughout the endpoint-dipole analysis we treat the direction of $\bm v_w$ as a fixed template.  In the standard scenario where the CNB rest frame coincides with the CMB rest frame, this direction is fixed by the measured CMB dipole. If local gravitational clustering or a halo-dependent neutrino bulk flow makes the CNB rest frame closer to the local matter frame than to the CMB frame, both the magnitude and direction of $\bm v_w$ become model dependent.  In such a case one would need to fit a dipole vector, rather than a single fixed angular template in $u=\hat{\bm p}_e\cdot\hat{\bm v}_w$. This would introduce additional angular parameters and can only weaken the sensitivity relative to the fixed-template estimates presented here, although the order-of-magnitude exposure scaling is unchanged.

We make two assumptions about the angular bins. First, the width of the angular bins is much larger than the angular resolution. We therefore neglect angular smearing. Second, we assume $\sum_{j=1}^{N_{\Omega}} u_j=0$, which corresponds to $\int_{-1}^{1}u\,\D u=0$ in the continuous limit. This condition is important for making the analysis robust against systematic uncertainties in the energy resolution, as we will discuss in Sec.~\ref{sec:statistics_energy_angle}.

The isotropic rate density per unit $u$ consists of the $\beta$-decay background and the isotropic CNB capture rate, uniformly distributed over $u\in[-1,1]$:
\begin{align}
    \Gamma_{\rm B}^{({\rm dipole}),X}\equiv \frac{1}{2}(\Gamma_{\rm B}^{({\rm rate})} + \Gamma_{\rm S}^{({\rm rate}),X})\,,
    \label{eq:gamma_B_Omega}
\end{align}
where $\Gamma_{\rm B}^{({\rm dipole}),X}$ denotes the isotropic rate density per unit $u=\cos\theta$.
In this dipole test, the isotropic CNB contribution is included in the isotropic component, while the signal template tested below is the dipole amplitude.

Note that $\Gamma_{\rm B}^{({\rm dipole}),X}\geq \Gamma_{\rm S}^{({\rm rate}),X}/2>0$. Therefore,
\begin{align}
    \frac{\Gamma_{\rm S}^{({\rm dipole}),X}}{\Gamma_{\rm B}^{({\rm dipole}),X}}\leq \frac{\Gamma_{\rm S}^{({\rm dipole}),X}}{\frac{1}{2}\Gamma_{\rm S}^{({\rm rate}),X}}=\frac{\delta\Gamma_{\rm CNB}}{\Gamma_{\rm CNB,0}}\leq1\,.
\end{align}
This is in contrast to the endpoint-rate analysis, where the corresponding ratio $r_{\rm S/B}^{({\rm rate}),X}$ can exceed unity when the effective resolution is sufficiently small, $\Delta E_{\rm eff}\lesssim 0.7m_\nu$.

The expected total count in the $j$-th angular bin is
\begin{align}
    \lambda^{({\rm dipole}),X}_j=(S^{({\rm dipole}),X}u_j+B^{({\rm dipole}),X})\Delta u
    \,,\label{eq:lambda_Omega}
\end{align}
where
$S^{({\rm dipole}),X}\equiv\Xi\Gamma_{\rm S}^{({\rm dipole}),X}/m_{\Tritium}$ and $B^{({\rm dipole}),X}\equiv \Xi\Gamma_{\rm B}^{({\rm dipole}),X}/m_{\Tritium}$ denote the anisotropic signal amplitude and the isotropic count density per unit $u$ for exposure $\Xi$, respectively.

We do not attempt a detector design study in this work; nevertheless, obtaining some angular information is not obviously impossible in principle. Because the signal enters as a dipolar modulation in $u=\cos\theta_{ew}$, the endpoint-window angular information can be compressed into a forward--backward (two-bin) observable without changing the qualitative scaling. More generally, any symmetric binning satisfying $\sum_j u_j\,\Delta u = 0$ (including the two-bin choice) preserves the cancellation of the overall normalization nuisance which will be discussed in Sec.~\ref{sec:statistics_energy_angle}. 

For PTOLEMY-type experiment, an endpoint-angular measurement has not yet been demonstrated, although CRES-based reconstruction of the parallel and transverse momentum components of endpoint electrons is under development\cite{Iwasaki:2024voi}. 
As a related reference, KATRIN has developed angular-selective photoelectron sources for MAC-E-filter calibration, demonstrating controlled electron injection with adjustable pitch angle and characterized angular spread \cite{Beck:2014xfa,Behrens:2017cmd,Schneidewind:2026ole}. 
These measurements calibrate the spectrometer response rather than the endpoint-bin dipole template itself, but they indicate that coarse angular information at the scale relevant for a forward–backward analysis is not an unreasonable idealization. We therefore neglect explicit angular smearing in the main analysis and encode residual angular-response uncertainties through nuisance parameters in Sec.~\ref{sec:statistics_energy_angle}.

It is useful to contrast our endpoint-dipole strategy with the polarized-target proposal of Ref.~\cite{Lisanti:2014pqa}.  In that approach the CNB anisotropy is accessed through spin-dependent terms in the capture cross section, requiring a macroscopic polarization of the tritium target and control of spin-dependent systematics.  Our strategy instead assumes an unpolarized target and attempts to extract the wind from the angular distribution of endpoint electrons. Thus, it trades the challenge of target polarization for the challenges of measuring coarse endpoint-electron angular information and accumulating a much larger exposure.  A detailed optimization of these two experimental strategies is detector dependent and beyond the scope of this work, but both should be viewed as complementary approaches to the same CNB-wind observable.

\section{Statistical framework for projected sensitivity}
\label{sec:statistics}

In this section we employ the profile likelihood method to derive expressions for the projected sensitivity of CNB searches in both the endpoint-rate analysis strategy and the endpoint-dipole analysis strategy, which were described in the previous section. 
Numerical results are presented in Sec.~\ref{sec:statistics_plot}.

\subsection{Profile likelihood and Asimov approximation}
\label{sec:framework}

In this work we adopt a profile-likelihood framework following Ref.~\cite{Cowan:2010js}.
We use a binned likelihood $\mathcal{L}(\{n_j\};\mu,\bm\theta)$ with signal strength parameter $\mu$ and
nuisance parameters $\bm\theta$. 
The nuisance parameters account for systematic uncertainties in the energy and angular distributions; they are discussed further in the following subsections.  
The signal strength parameter is introduced by rescaling the endpoint signal templates, $S^{({\rm rate}),X}$ and $S^{({\rm dipole}),X}$ from Sec.~\ref{sec:observables}, as $S \to \mu S$.
For observed counts $\{n_j\}$ and expected counts $\{\lambda_j(\mu,\bm\theta)\}$, the likelihood is taken to be 
\begin{align}\label{eq:likelihood}
    \mathcal{L}(\{n_j\};\mu,\bm\theta)
    =
    \prod_j
    {\rm Pois}\!\left(n_j;\lambda_j(\mu,\bm\theta)\right)
    \prod_a
    \mathcal{N}\!\left(\theta_a;0,\sigma_a^2\right)\,.
\end{align}
Here ${\rm Pois}(n;\lambda)$ is the Poisson probability mass function with mean $\lambda$, while $\mathcal{N}(\theta_a;0,\sigma_a^2)$ is a Gaussian prior on the nuisance parameter $\theta_a$ with mean zero and variance $\sigma_a^2$.

We estimate the median sensitivity $Z$ by testing $\mu=0$ against $\mu>0$, using the Asimov data set $n_j^A=\lambda_j(\mu_{\rm true},\bm\theta_0)$ generated at $\mu_{\rm true}=1$ with nominal nuisance parameters $\bm\theta_0=\bm 0$. 
In the large-sample limit, the profiled uncertainty $\sigma_\mu$ is calculated using the inverse Fisher matrix as
\begin{align}
    \sigma_\mu^2
    =
    \left(I^{-1}\right)_{\mu\mu},
    \qquad
    I_{\alpha\beta}
    =
    -\left.
    \frac{\partial^2 \ln \mathcal{L}(\{n_j^A\};\bm\eta)}
    {\partial \eta_\alpha \partial \eta_\beta}
    \right|_{\bm\eta=\bm\eta_0},
    \qquad
    \bm\eta=(\mu,\bm\theta)\,.
\end{align}
Here the inverse is taken after including the nuisance parameters, which
implements profiling over them. 
The median significance is then estimated as 
\begin{align}\label{eq:Zmed}
    Z
    \simeq
    \frac{\mu_{\rm true}}{\sigma_\mu}\,.
\end{align}
In the following, we take $\mu_{\rm true} = 1$ and evaluate $\sigma_\mu$ for the endpoint-rate and endpoint-dipole analyses. 
We assess the requirements for a $z$-$\sigma$ discovery of the CNB signal (e.g., $z = 3$) by imposing $Z \geq z$.

\subsection{Endpoint-rate analysis strategy}
\label{sec:statistics_energy_only}

In the endpoint-rate analysis strategy, the observable is the event rate $\lambda^{{\rm rate},X}$ in the endpoint region; see Sec.~\ref{sec:SB in energy_only}. 
To employ the profile likelihood framework, we introduce a signal strength parameter $\mu$ and a single nuisance parameter $\alpha$, to express the event rate as 
\begin{align}
    \lambda^{({\rm rate}),X}_{\rm sys}(\mu,\alpha)\equiv (1+\alpha)\left[\mu S^{({\rm rate}),X} + B^{({\rm rate})}\right]\,.
    \label{eq:lambda_sys_E}
\end{align}
The nuisance parameter $\alpha$ scales both the signal and background; it is associated with, for example, imperfect modeling of the effective energy response, acceptance, final-state effects, or subleading corrections to the $\beta$-decay spectrum. 
As indicated by the Gaussian likelihood in \Eref{eq:likelihood}, we assume that $\alpha$ has mean zero and variance $\sigma_\alpha^2$.  
For a PTOLEMY-like tritium-on-graphene setup, Ref.~\cite{Cheipesh:2021fmg} estimated the effective endpoint broadening for several models of the tritium trapping potential and found that the resulting response can vary at least at the $O(10\%)$ level; this motivates $\sigma_\alpha\simeq 10^{-1}$ as a conservative phenomenological benchmark. As a more optimistic reference, the KATRIN windowless gaseous tritium source requires the column density and source activity to be stabilized and monitored at the 0.1\% level, and its first tritium commissioning run demonstrated stable operating conditions at this level \cite{Babutzka:2012xd}, suggesting $\sigma_\alpha\simeq 10^{-3}$ as an experimentally achieved normalization-control scale.

In the background-dominated regime, $S^{({\rm rate}),X} \ll B^{({\rm rate})}$, we calculate the approximate median significance using \Eref{eq:Zmed} to find 
\begin{align}
    Z^{({\rm rate}),X}\simeq \frac{S^{({\rm rate}),X}}{\sqrt{B^{({\rm rate})} + (B^{({\rm rate})}\sigma_\alpha)^{2}}}= \frac{\Xi\Gamma_{\rm S}^{({\rm rate}),X}/m_{\Tritium}}{\sqrt{\Xi\Gamma_{\rm B}^{({\rm rate})}/m_{\Tritium} + (\Xi\Gamma_{\rm B}^{({\rm rate})}\sigma_\alpha/m_{\Tritium})^{2}}}\,.
\end{align}
If the exposure $\Xi$ is small, then $Z^{(\mathrm{rate}),X} \propto \Xi^{1/2}$ and the sensitivity is statistics limited.  
Reaching a $3\sigma$ discovery of the CNB signal requires a sufficiently large exposure 
\begin{align}
    Z^{({\rm rate}),X} \ge 3 
    \quad \Rightarrow \quad 
    \Xi\gtrsim \Xi_{\rm req}^{({\rm rate}),X} 
    \equiv \frac{9 \Gamma_{\rm B}^{({\rm rate})} \, m_{\Tritium}}{\bigl( \Gamma_{\rm S}^{({\rm rate}),X} \bigr)^{2}}\,.
    \label{eq:Xi_required_E}
\end{align}
If the exposure $\Xi$ is large, then $Z_\mathrm{med}^{(\mathrm{rate}),X} \propto \Xi^{0}$ and the sensitivity is systematics limited.  
Reaching a $3\sigma$ discovery of the CNB signal requires a sufficiently small uncertainty on the nuisance parameter 
\begin{align}
    Z^{({\rm rate}),X} \ge 3 
    \quad \Rightarrow \quad 
    \sigma_{\alpha}\lesssim (\sigma_{\alpha})^{X}_{\rm req} \equiv \frac{\Gamma_{\rm S}^{({\rm rate}),X}}{3 \Gamma_{\rm B}^{({\rm rate})}}\,.
    \label{eq:sigma_alpha_required}
\end{align}
We plot these quantities and provide numerical results in Sec.~\ref{sec:results}. 

\subsection{Endpoint-dipole analysis strategy}
\label{sec:statistics_energy_angle}

In the endpoint-dipole analysis strategy, the observables are the event rates $\lambda_j^{{\rm dipole},X}$ across angular bins at the endpoint region; see Sec.~\ref{sec:SB in energy_angle}. 
To employ the profile likelihood framework, we introduce a signal strength parameter $\mu$ and three nuisance parameters $\alpha$, $\kappa$, and $\zeta$, to express the event rate as 
\begin{align}
    {\lambda_j}^{({\rm dipole}),X}_{\rm sys}(\mu,\alpha,\kappa,\zeta)\equiv (1+\alpha)\left[\mu (1+\kappa)S^{({\rm dipole}),X}u_j + \zeta B^{({\rm dipole}),X}u_j + B^{({\rm dipole}),X}\right]\Delta u\,.
    \label{eq:lambda_sys_Omega}
\end{align}
Recall that $u_j$ denotes the bin-averaged value of $u = \cos\theta_{ew} = \hat{\bm p}_e \cdot \hat{\bm v}_w$ in the $j$-th angular bin, $N_{\Omega}$ is the number of angular bins, $j \in \{ 1, \cdots, N_{\Omega} \}$, and $\Delta u = 2 / N_{\Omega}$ is the width of an angular bin. 
The nuisance parameter $\alpha$ scales the signal and background; it has the same interpretation that we discussed in Sec.~\ref{sec:statistics_energy_only}.  
The nuisance parameter $\kappa$ scales the signal; it is associated with, for example, a multiplicative uncertainty in the dipole response, angular acceptance/resolution, or dipole calibration.  
The nuisance parameter $\zeta$ scales the dipolar component of the background; it represents the residual fractional dipole component of the otherwise isotropic endpoint background after calibration of the angular acceptance and electron transport.  For example, a geometric or transport asymmetry that makes the beta-decay endpoint background appear proportional to $1+\zeta u$ would generate precisely the term $\zeta B^{({\rm dipole}),X}u_j$ in \Eref{eq:lambda_sys_Omega}.
\footnote{
Since the $v_w$-induced anisotropic CNB signal enters as a dipolar modulation in the laboratory frame, the leading instrumental/background anisotropy that can mimic it is likewise well captured by a dipole term; higher multipoles are subleading for the specific test we consider.}

As indicated by the Gaussian likelihood in \Eref{eq:likelihood}, we assume that each of the nuisance parameters have mean zero, and we denote their variances by $\sigma_\alpha^2$, $\sigma_\kappa^2$, and $\sigma_\zeta^2$. 
Since no endpoint-dipole measurement has yet been performed, existing experiments do not provide direct measurements of $\sigma_\kappa$ $\sigma_\zeta$; in the following we therefore treat them as phenomenological control parameters.  As a reference scale, we use $\sigma_\kappa=0.1$ for residual multiplicative angular-response uncertainties, 
while leaving $\sigma_\zeta$ to be constrained by the sensitivity requirement and, later, by sideband data.

In the background-dominated regime, $S^{({\rm dipole}),X} \ll B^{({\rm dipole})}$, we calculate the approximate median significance using \Eref{eq:Zmed} to find 
\begin{align}
    Z^{({\rm dipole}),X}
    & \simeq \frac{S^{({\rm dipole}),X}}{\sqrt{C^{-1}B^{({\rm dipole}),X} +(B^{({\rm dipole}),X}\sigma_\zeta)^{2}+(S^{({\rm dipole}),X}\sigma_\kappa)^{2}}} 
    \nonumber\\
    & = \frac{\Xi\Gamma_{\rm S}^{({\rm dipole}),X}/m_{\Tritium}}{\sqrt{C^{-1}\Xi\Gamma_{\rm B}^{({\rm dipole}),X}/m_{\Tritium} +(\Xi\Gamma_{\rm B}^{({\rm dipole}),X}\sigma_\zeta/m_{\Tritium})^{2}+(\Xi\Gamma_{\rm S}^{({\rm dipole}),X}\sigma_\kappa/m_{\Tritium})^{2}}} \,,
    \label{eq:Z_med_Omega}
\end{align}
with $C \equiv \sum_j u_j^2 \, \Delta u$. 
If the exposure $\Xi$ is small, then $Z^{(\mathrm{dipole}),X} \propto \Xi^{1/2}$ and the sensitivity is statistics limited.  
Reaching a $3\sigma$ discovery of the CNB signal requires a sufficiently large exposure 
\begin{align}
    Z^{({\rm dipole}),X} \ge 3 
    \quad \Rightarrow \quad 
    \Xi 
    \gtrsim \Xi_{\rm req}^{({\rm dipole}),X} 
    \equiv \frac{9 \Gamma_{\rm B}^{({\rm dipole}),X} \, m_{\Tritium}}{C\bigl( \Gamma_{\rm S}^{({\rm dipole}),X} \bigr)^{2}} 
    \,.
    \label{eq:Xi_req_Omega}
\end{align}
If the exposure $\Xi$ is large, then $Z^{(\mathrm{dipole}),X} \propto \Xi^{0}$ and the sensitivity is systematics limited.  
Reaching a $3\sigma$ discovery of the CNB signal requires a sufficiently small uncertainty on the nuisance parameters 
\begin{align}\label{eq:sigma_kappa_req}
    Z^{({\rm dipole}),X} \ge 3 
    \quad \Rightarrow \quad 
    \begin{cases}
    \sigma_\zeta\lesssim (\sigma_\zeta)_{\rm req}^{X} \equiv \frac{\Gamma_{\rm S}^{({\rm dipole}),X}}{3\Gamma_{\rm B}^{({\rm dipole}),X}}
    \\ 
    \sigma_{\kappa}\lesssim (\sigma_{\kappa})_{\rm req} \equiv \frac{1}{3}
    \end{cases}\,.
\end{align}

The constraint on $\sigma_\zeta$ in \Eref{eq:sigma_kappa_req} has a simple interpretation.
The fake-dipole term in \Eref{eq:lambda_sys_Omega} is proportional to $\zeta B^{({\rm dipole}),X}u_j$, while the CNB-wind signal template is proportional to $S^{({\rm dipole}),X}u_j$.  Therefore, a residual background dipole with
\begin{align}
    |\zeta| \sim
  \frac{S^{({\rm dipole}),X}}{B^{({\rm dipole}),X}}
  =
  \frac{\Gamma_{\rm S}^{({\rm dipole}),X}}
       {\Gamma_{\rm B}^{({\rm dipole}),X}}
\end{align}
would mimic the CNB-induced dipole at the level of the signal being tested. A $3\sigma$ discovery requires the uncertainty on such a fake dipole to be smaller by approximately a factor of three, giving the first condition in \Eref{eq:sigma_kappa_req}.
In the small (i.e~good) energy resolution regime $\Delta E_{\rm eff}\lesssim 0.7m_\nu$, we have $\Gamma_{\rm B}\sim \Gamma_{\rm CNB,0}^{X}/2$. Thus, for nonrelativistic Dirac neutrinos with $m_\nu\simeq0.1\,{\rm eV}$, the residual dipolar component of the endpoint-background acceptance must be controlled at the $10^{-3}$ level.  For Majorana neutrinos the corresponding estimate is at the $10^{-6}$ level.  In the large (poor) energy resolution regime, the requirement is further tightened by the endpoint signal-to-background ratio.

It is interesting to note that the significance \eqref{eq:Z_med_Omega} does not depend on $\sigma_\alpha$, which is the uncertainty in the nuisance parameter $\alpha$.  
In fact, $\alpha$ is orthogonal to the other parameters in the Fisher matrix in the sense that $I_{\mu\alpha} = I_{\kappa\alpha} = I_{\zeta\alpha} = 0$.  
For example, 
\begin{align}
    I_{\mu\alpha} 
    = \sum_j\frac{1}{{\lambda_j}^{({\rm dipole}),X}_{\rm sys}|_{\bm{\eta}=\bm{\eta}_0}}\left(S^{({\rm dipole}),X}u_j\Delta u\right)\left({\lambda_j}^{({\rm dipole}),X}_{\rm sys}|_{\bm{\eta}=\bm{\eta}_0}\right) 
    = S^{({\rm dipole}),X} \, \sum_j u_j \Delta u = 0\,.
    \label{eq:fisher matrix}
\end{align}
In the last equality we used $\sum_j u_j \Delta u = 0$, which assumes a symmetric angular binning. 
For symmetric binning, the normalization nuisance parameter $\alpha$ changes the total number of events equally in all angular bins and therefore cannot mimic the angular dipole signal.  
As a result, $\alpha$ does not affect the discovery potential.
It is worth emphasizing that both the endpoint-rate and the endpoint-dipole analysis strategies are systematics limited when the exposure is large, but they are controlled by different systematics: $\alpha$ for endpoint-rate and either $\kappa$ or $\zeta$ for endpoint-dipole.

As a final remark, we note that unlike $\alpha$ in the endpoint-rate analysis, the fake dipole
parameter $\zeta$ can in principle be constrained by using energy sidebands that contain no CNB signal, i.e.,~an off-bin (sideband) region sufficiently below the CNB line. The mean counts satisfy,
\begin{align}
    {\lambda_j}_{\rm sys}^{(\rm off)}\simeq (1+\alpha)B^{(\rm off)}(1+\zeta u_j)\Delta u\,,
    \label{eq:lambda_j_off}
\end{align}
where $B^{(\rm off)}=\Xi \Gamma^{(\rm off)}_{\rm B}/2m_{\Tritium}$ is the isotropic background yield in the sideband from $\beta$-decay,\footnote{
The factor $1/2$ converts the sideband yield integrated over $u\in[-1,1]$ into a density per unit $u$, in the same way as $\Gamma_{\rm B}^{({\rm dipole}),X}$ is defined in \Eref{eq:gamma_B_Omega}.
}
\begin{align}
    \Gamma^{(\rm off)}_{\rm B}\equiv \int _{L^{(\rm off)}}^{U^{(\rm off)}} \D E_{e}\int \D \Omega_e\,\frac{\D ^{2}\tilde{\Gamma}_{\beta}}{\D E_{e}\D \Omega_e}\,.
\end{align}
For example, we consider a single energy bin centered at $E_{\rm pk}-2\Delta E_{\rm eff}$, i.e.,~$L^{(\rm off)}=E_{\rm pk}-5\Delta E_{\rm eff}/2$ and $U^{(\rm off)}=E_{\rm pk}-3\Delta E_{\rm eff}/2$. It corresponds to $B^{\rm (off)}=3.5\times 10^8(\Xi/100\,{\rm g\cdot yr})$ for $\Delta E_{\rm eff}=m_\nu=0.1\,{\rm eV}$.
The Fisher information for $\zeta$ extracted from the off-bin scales as $I^{(\rm off)}_{\zeta \zeta}\sim CB^{(\rm off)}$, implying
\begin{align}
    \sigma_\zeta^{({\rm off})}\simeq \sqrt{\frac{2}{CB^{(\rm off)}}}\,.
\end{align}
Using \eqref{eq:Xi_req_Omega} and $B^{(\rm off)}\geq B^{({\rm dipole}),X}$, we obtain
\begin{align}
    \sigma_\zeta^{({\rm off})}\lesssim \frac{\Gamma_{\rm S}^{({\rm dipole}),X}}{z\Gamma_{\rm B}^{({\rm dipole}),X}}=(\sigma_\zeta)_{\rm req}^{X}(z)\,.
\end{align}
Therefore, if the fake dipole is indeed approximately energy independent over
the on/off-bin separation, we may have a natural handle to constrain $\zeta$ with high statistics since
$B^{({\rm off})}\geq B^{({\rm dipole}),X}$.

\section{Results}
\label{sec:results}
\label{sec:statistics_plot}

How much exposure $\Xi$ would be required to detect the CNB through neutrino capture on tritium, and how much additional exposure would be required to reveal the dipolar anisotropy of the CNB wind?  
The answers depend on properties of the CNB such as the mass scale $m_\nu$ and the Dirac/Majorana nature of these particles.  
The answers also depend on the experiment's energy resolution $\Delta E_{\rm eff}$ and on the experimentalist's ability to reduce systematic uncertainties $\sigma_\alpha$.  
In this section we present two figures, which summarize our main results.  

\subsection{Required exposure}
\label{sec:required exposure}

\begin{figure}[t]
  \centering
  \begin{subfigure}{0.48\textwidth}
    \centering
    \includegraphics[width=\linewidth]{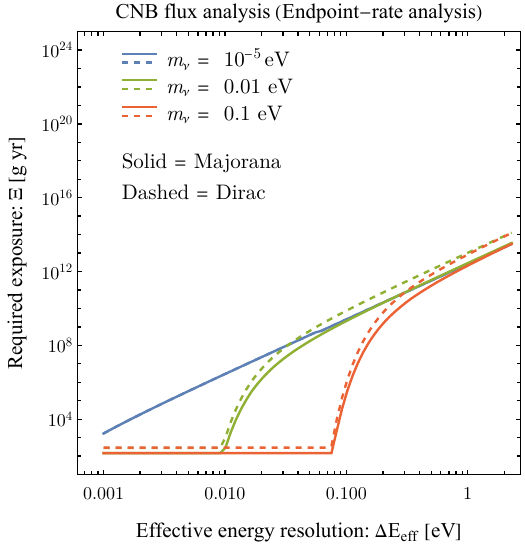}
    \label{fig:exampleA}
  \end{subfigure}
  \hfill
  \begin{subfigure}{0.48\textwidth}
    \centering
    \includegraphics[width=\linewidth]{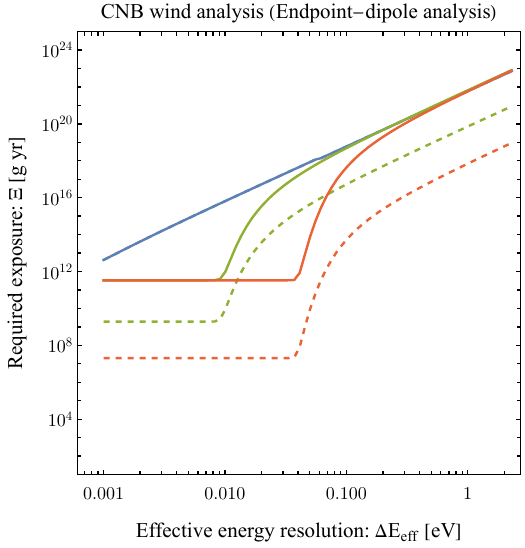}
    \label{fig:exampleB}
  \end{subfigure}
    \caption{
    \label{fig:required exposure}
    The minimum exposure $\Xi$ required for CNB searches as a function of the effective energy resolution $\Delta E_{\rm eff}$.  
    \textit{Left:} Our estimate of the required exposure for a $3\sigma$ detection of the CNB flux via the endpoint-rate analysis strategy is given by \Eref{eq:Xi_required_E}. 
    \textit{Right:} Our estimate of the required exposure for a $3\sigma$ detection of the CNB wind via the endpoint-dipole analysis strategy is given by \Eref{eq:Xi_req_Omega}. 
    }
\end{figure}

Figure \ref{fig:required exposure} shows the minimum exposure required for a $3\sigma$ detection of the CNB as a function of the effective energy resolution $\Delta E_{\rm eff}$.  
We separately show the requirement for detecting the isotropic CNB flux and for detecting the dipolar CNB wind.  
The left panel corresponds to the endpoint-rate analysis strategy, whose statistics-limited exposure requirement is given in \Eref{eq:Xi_required_E}.  
The right panel corresponds to the endpoint-dipole analysis strategy, whose corresponding requirement is given in \Eref{eq:Xi_req_Omega}.  
In both panels the exposure is quoted in units of $\rm g \cdot yr$, 
and we show three representative neutrino masses, $m_\nu=10^{-5}\,{\rm eV}, 0.01\,{\rm eV}$ and $0.1\,{\rm eV}$, for both Majorana and Dirac neutrinos.

The smallest mass value, $m_\nu=10^{-5}\,{\rm eV}$, should be understood as an effectively massless benchmark.
\footnote{
We have also checked that replacing the effectively massless benchmark by the
limiting normal- or inverted-ordering spectra, with one effectively massless eigenstate and the other masses fixed by the measured mass splittings, does not
qualitatively change the curves shown in Fig.~\ref{fig:required exposure}.  We therefore do not show these additional benchmarks separately.
} 
It is included to illustrate the limiting regime $m_\nu\ll \Delta E_{\rm eff}$, where the endpoint-rate sensitivity becomes insensitive to the neutrino mass scale.  It should not be interpreted as the physical normal-hierarchy spectrum with $m_{\rm lightest}=10^{-5}\,{\rm eV}$. A realistic hierarchical spectrum would require an explicit sum over the three physical masses and PMNS weights.
Since this benchmark also satisfies \(m_\nu\ll T_\nu\), the relic neutrinos are relativistic.  In this limit the Dirac and Majorana rates become nearly identical, because \(A^D=1+v_\nu\simeq 2=A^M\) for the isotropic component and \(B^D=1+v_\nu\simeq 2\simeq B^M=2v_\nu\) for the dipole component.

The left panel of Fig.~\ref{fig:required exposure} corresponds to the familiar endpoint-rate analysis strategy, which searches for the CNB flux by measuring the electron recoil rate at the endpoint of the beta decay spectrum. 
Along the horizontal branches of the orange and green curves, the effective energy resolution has $\Delta E_{\rm eff} \lesssim 0.7 m_\nu$.
With this relatively small (good) energy resolution, measurements of the beta decay spectrum could relatively easily distinguish the signal (monochromatic CNB capture line at $m_\nu$ above the endpoint) from the background (smeared beta-decay tail from below the endpoint).  
Since the signal rate exceeds the background rate, the required exposure for $3\sigma$ discovery is comparatively modest: $\Xi \gtrsim 290\,{\rm g\,yr}$ for Dirac neutrinos and $145\,{\rm g\,yr}$ for Majorana, which agrees with earlier work \cite{Long:2014zva, Betts:2013uya}.  
Along the diagonal branch of each curve, the effective energy resolution has $\Delta E_{\rm eff} \gtrsim \mathrm{few} \times m_\nu$. 
With this relatively large (poor) energy resolution, the beta decay background electrons would contaminate the CNB signal region above the endpoint. 
Since the background rate exceeds the signal rate, discovery of the CNB flux would require a large exposure $\Xi \propto (\Delta E_{\rm eff})^{3}$ approximately 
and an excellent control of systematic uncertainties (see Fig.~\ref{fig:required systematics}, discussed below). 

It is interesting to compare how the exposure requirement differs for Dirac and Majorana neutrinos.  
The isotropic component of the capture rate \eqref{eq:Gamma_CNB} depends on the Dirac/Majorana nature of the neutrinos via the spin-dependent factor $\mathcal{I}_A^X \propto A^{X}$. 
If the neutrinos are Majorana, then both positive- and negative-helicity particles contribute giving $A^M = 2$, but if the neutrinos are Dirac, then only negative-helicity particles contribute giving $A^D = 1 + v_\nu$.  
For nonrelativistic neutrinos $v_\nu \approx T_\nu / m_\nu \ll 1$, this leads to a factor of $A^M / A^D \approx 2$ difference in the signal rates, and corresponding factor of $2$ difference in the required exposure (cf., red-solid and red-dashed). 
For relativistic neutrinos $v_\nu \approx 1$, this leads to a factor of $A^M / A^D \approx 1$ difference in the signal rates, and similar values for the required exposures (cf., blue-solid and blue-dashed).  

The right panel of Fig.~\ref{fig:required exposure} corresponds to our endpoint-dipole analysis strategy, which searches for the CNB wind by measuring the electron recoil rate across a range of angular bins at the endpoint of the beta decay spectrum. 
Along the horizontal branches of the orange and green curves, the energy resolution is relatively small, and the signal rate exceeds the background rate.  
However, the signal rate for the CNB wind search (dipole component) is much smaller than the signal rate for the CNB flux search (isotropic component), and a commensurately larger exposure $\Xi$ is required to achieve a $3\sigma$ detection. 
For Majorana neutrinos (solid curves), the required exposure is insensitive to the mass for $\Delta E_{\rm eff} \lesssim 0.7 m_\nu$, and it evaluates to $\Xi \gtrsim 3 \times 10^{11} \, {\rm g\,yr}$.  
For Dirac neutrinos (dashed curves), the signal rate is enhanced compared to Majorana by the ratio of speeds $v_w / \bar{v}_\nu = v_w m_\nu / T_\nu$, see \Eref{eq:RBA scaling}, and the required exposure is relatively much smaller, albeit still large in absolute terms.  
Along the diagonal branch of each curve, the effective energy resolution exceeds the neutrino mass scale, and the background rate exceeds the signal rate, requiring an even larger exposure to achieve a $3\sigma$ detection of the CNB wind.  
Needless to say, the enormous exposures shown on these plots are experimentally impractical, but they are a convenience reference scale by which we can quantify the difficulty of detecting the CNB flux, and the vastly more challenging task of detecting the CNB wind.  

\subsection{Required systematic uncertainty}
\label{sec:required systematics}

\begin{figure}[t]
    \centering
    \includegraphics[width=0.70\linewidth]{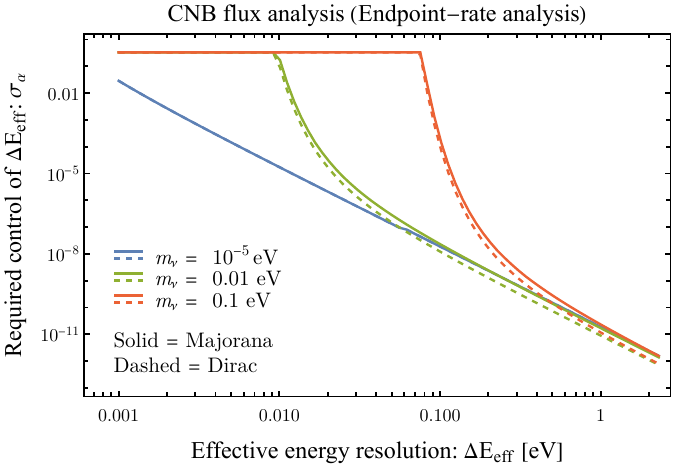}
    \caption{
        The maximum systematic uncertainty $\sigma_\alpha$ allowed for a $3\sigma$ detection of the CNB flux in the endpoint-rate analysis strategy.  
        The nuisance parameter $\alpha$, which rescales the signal and background, has a Gaussian prior with variance $\sigma_\alpha^2$. 
        Color and dashing are the same as in Fig.~\ref{fig:required exposure}.
    } 
    \label{fig:required systematics}
\end{figure}

Figure \ref{fig:required systematics} shows the maximum systematic uncertainty allowed for a $3\sigma$ detection of the CNB as a function of the effective energy resolution $\Delta E_{\rm eff}$.  
In the endpoint-rate analysis strategy, the selected endpoint bin contains both the CNB capture signal and the smeared beta-decay background.  
We model residual uncertainty in the predicted endpoint-bin normalization by the dimensionless nuisance parameter $\alpha$, as in \Eref{eq:lambda_sys_E}.  
We ascribe $\alpha$ a Gaussian prior with variance $\sigma_\alpha^2$, and the maximum allowed value for a $3\sigma$ detection is given in \Eref{eq:sigma_alpha_required}.

The main feature of Fig.~\ref{fig:required systematics} follows directly from the signal-to-background ratio in the endpoint bin.  
For $\Delta E_{\rm eff} \lesssim m_\nu$, the capture peak is effectively resolved from the beta-decay tail, the endpoint-rate analysis is signal dominated, and the allowed $\sigma_\alpha$ can be $1/3$ for $3\sigma$ discovery. 
By contrast, for $\Delta E_{\rm eff} \gtrsim m_\nu$, the beta-decay background contaminates the signal region.  
In this regime the CNB signal must be extracted as a small excess over a much larger background, and the required control of $\sigma_\alpha$ becomes very stringent.

This behavior is especially important for realistic energy resolutions. Around $\Delta E_{\rm eff}\sim 0.1\,{\rm eV}$, the required value of $\sigma_\alpha$ can already be far below the percent level, depending on the neutrino mass. At poorer resolutions the requirement becomes even tighter. Thus, in the endpoint-rate analysis, exposure alone is not sufficient: the endpoint response, final-state effects, and residual corrections to the beta-decay spectrum must also be controlled with very high precision.

In closing we remark that the endpoint-dipole analysis strategy is affected by systematics in a different way than the endpoint-rate analysis strategy. 
Because the dipole template is odd in $u=\cos\theta$, while the dominant endpoint-normalization uncertainty is isotropic, the nuisance parameter $\alpha$ cancels for symmetric angular binning, as shown in \eqref{eq:fisher matrix}. The relevant residual systematics are instead the fake dipole contribution $\zeta$ and the dipole-response calibration $\kappa$, introduced in \eqref{eq:lambda_sys_Omega}. Their impact is summarized by \Eref{eq:sigma_kappa_req}. In particular, an additive fake dipole can in principle be constrained using sideband data, while the multiplicative calibration uncertainty only produces a limiting floor when $\sigma_\kappa\gtrsim 1/z$. Therefore, the CNB wind search trades a much smaller signal amplitude for reduced sensitivity to the endpoint-normalization uncertainty.

\section{Conclusion}
\label{sec:conclusions}

In this work we studied how the capture of CNB relic neutrinos on tritium would result in a dipolar anisotropy in the momentum of the recoiling electrons, and we assessed the experimental parameters that would be required to detect such a signal.  
This work serves to quantify the difficulty of detecting the CNB wind.  
In this section we briefly summarize our work, and offer a few remarks on outlook.  

On the theoretical side, we calculated the anisotropic signal rate.  
To do so we assume a $\Lambda$CDM cosmology, which predicts that the CNB neutrinos have an isotropic thermal momentum distribution with effective temperature $T_\nu \approx 0.168 \; \mathrm{meV}$ and that the CNB rest frame coincides with the CMB rest frame.  
A laboratory on Earth moves relative to the CNB rest frame at a typical speed of $v_w \sim 10^{-3}$, leading to an approximately dipolar anisotropy in the flux of incident neutrinos, which is called the CNB wind.  
We calculate the differential rate for relic neutrino capture on an unpolarized tritium nucleus through the channel $\nu_i + \Tritium \to \He + e^{-}$ and we show that the recoiling electrons are distributed with a dipolar anisotropy.  
We find that although the CNB flux anisotropy can be large, $\delta\Phi_\nu / \Phi_{\nu,0} \sim v_w / v_\nu \sim (m_\nu / 100 \; \mathrm{meV})$, the electron emission anisotropy is generally small, $\delta\Gamma_{\rm CNB} / \Gamma_{{\rm CNB},0} \sim 10^{-3}$ for nonrelativistic Dirac neutrinos with $m_\nu\sim 0.1 \, {\rm eV}$, and $\sim 10^{-6}$ for Majorana neutrinos. 
This suppression arises from the small outgoing-electron velocity, $v_e/c\ll 1$, together with a partial cancellation in the nuclear form-factor combination that controls the electron–neutrino angular correlation. 
One can see that a large exposure would be required to detect the CNB wind, even in the absence of backgrounds and experimental noise.  

On the experimental side, we model the background and noise, which generally dominates over the signal, and we propose a strategy for revealing the CNB wind signal.  
The dominant background is expected to be the tritium beta decay itself, which contaminates the signal region above the endpoint when the experimental energy resolution is not small compared to the neutrino mass scale.  
We consider a hypothetical experiment with a poor energy resolution, such that the expected number of background events in the endpoint region dwarfs the expected number of signal events.  
However, assuming that the background electrons are isotropically distributed in the lab frame, we leverage the anisotropy of the signal electrons that arise from the CNB wind.  
We estimate the enormous exposure and the excellent control of systematic uncertainties that would be required for a $3\sigma$ discovery.

Our main results are presented in Figs.~\ref{fig:required exposure}~and~\ref{fig:required systematics}.  
In Fig.~\ref{fig:required exposure} we show the minimum exposure $\Xi$ required to achieve a $3\sigma$ detection of the CNB wind under different assumptions about the experimental energy resolution $\Delta E_{\rm eff}$, the neutrino mass scale, and the Dirac/Majorana nature.  
For comparison we also show the exposure required to simply detect the CNB flux, but not the dipolar anisotropy induced by the CNB wind.  
In Fig.~\ref{fig:required systematics} we show the maximum systematic uncertainty that is allowed for CNB detection.  
In the signal-dominated regime where $\Delta E_{\rm eff} \lesssim 0.08 \; \mathrm{eV}$, we find that detecting the CNB flux would require $\Xi \sim 100 \; \mathrm{g} \, \mathrm{yr}$ in agreement with earlier work, whereas detecting the CNB wind would require $\sim 10^5$ times larger exposure.  
In the background-dominated regime where $\Delta E_{\rm eff} \gtrsim 0.1 \; \mathrm{eV}$, we find that extremely good control of systematic uncertainties is needed to allow for signal/background discrimination.

This work reflects our desire to quantify the challenges posed by CNB direct detection.  
We adopt exposure $\Xi$ as our primary measure of difficulty.  
Needless to say, an experiment that seeks to detect the CNB wind using the methods we have described here would encounter numerous other difficulties:  where to store the immense tritium target, how to simultaneously measure the recoiling electron's energy and direction, how to ensure that electrons reach the detector without scattering on the tritium, how to control for backgrounds (e.g., cosmic rays), and so on.  
Nevertheless, exposure $\Xi$ provides a simple and intuitive measure that allows us to quantify the difficulty of detecting the CNB wind.  
Moreover, we have focused on the CNB wind, since this seems to be the most accessible of the CNB's properties.  
It would be interesting to consider the various other CNB properties that were enumerated in the introduction and to ascribe a measure of difficulty to each.

\section*{Acknowledgments}
We thank Kevin J.~Kelly for helpful comments.
This material is based upon work supported (in part: A.J.L.) by the National Science Foundation under Grant No.~PHY-2412797. 
M.~U.~and M.~Y.~are supported by IBS under the project code, IBS-R018-D3. 
M.~Y.~is also supported by JSPS Grant-in-Aid for Scientific Research Number JP23K20843.

\appendix

\section{Neutrino distribution function}
\label{sec:derivation of distribution function}


Before decoupling $(T\gg T_{\rm dec}\sim {\rm MeV})$, neutrinos are in thermal equilibrium. In the CNB isotropic rest frame, the distribution with a finite mass $m_\nu$ is given by
\begin{align}
    f_{\rm CNB}(p_{\rm CNB};T\gg T_{\rm dec})=\frac{1}{\exp(\sqrt{p_{\rm CNB}^{2}+m_\nu^2}/T)+1}\,,
\end{align}
where $p_{\rm CNB}$ denotes the magnitude of the neutrino three-momentum in the CNB rest frame.

For simplicity, we assume instantaneous decoupling at $T=T_{\rm dec}$, with scale factor $a(T_{\rm dec})=a_{\rm dec}$.
Under this assumption, the collision term is negligible after decoupling, and the distribution function is conserved along free-streaming trajectories in phase space. Thus,
\begin{align}
    f_{\rm CNB}(p_{\rm CNB})=f_{\rm dec}(p_{\rm dec})
\end{align}
where
\begin{align}
    f_{\rm dec}(p_{\rm dec})\equiv \frac{1}{\exp(\sqrt{p_{\rm dec}^{2}+m_\nu^2}/T_{\rm dec})+1}\,.
\end{align}
Here $p_{\rm dec}$ denotes the magnitude of the three-momentum of neutrino at decoupling. We define the neutrino temperature parameter $T_\nu\equiv \frac{a_{\rm dec}}{a}T_{\rm dec}$; after $e^\pm$ annihilation it is related to the photon temperature by $T_\nu\simeq (4/11)^{1/3}T_{\rm CMB}$. 

With momentum redshift $p_{\rm CNB}=p_{\rm dec}a_{\rm dec}/a$, we obtain
\begin{align}
    f_{\rm CNB}(p_{\rm CNB})
    =\frac{1}{\exp(\sqrt{p_{\rm CNB}^{2}a^{2}/a^{2}_{\rm dec}+m_\nu^2}/T_{\rm dec})+1}
    =\frac{1}{\exp(\sqrt{p_{\rm CNB}^{2}+m_\nu^2 a^{2}_{\rm dec}/a^{2}}/T_{\nu})+1}\,.
\end{align}
The final form shows that the mass enters as an ``effective'' redshifted mass $m_\nu a_{\rm dec}/a$. We expand this form in terms of $\epsilon\equiv\left(\frac{m_\nu a_{\rm dec}}{a p_{\rm CNB}}\right)^2$ to recover \eqref{eq:fCNB} up to $O(\epsilon)$. 
For typical momenta $p_{\rm dec}\sim T_{\rm dec}\sim{\rm MeV}$, this gives
\begin{align}
    \epsilon
    =
    \left(\frac{m_\nu a_{\rm dec}}{a p_{\rm CNB}}\right)^2
    =
    \left(\frac{m_\nu}{p_{\rm dec}}\right)^2
    \sim
    10^{-14}
    \left(\frac{m_\nu}{0.1\,{\rm eV}}\right)^2
    \left(\frac{{\rm MeV}}{T_{\rm dec}}\right)^2 .
\end{align}


We now relate the CNB-frame momentum to the momentum measured in the laboratory frame. In the remainder of this appendix we denote laboratory-frame quantities by the subscript ``LAB''. This notation corresponds to the main-text notation through
\begin{align}
    p_{\rm LAB}=p_\nu,\qquad E_{\rm LAB}=E_\nu ,
\end{align}
where $p_\nu$ and $E_\nu$ are the neutrino momentum magnitude and energy in the tritium rest frame used in Sec.~\ref{sec:signal}.

With the CNB rest frame moving with velocity $\bm v_w$ with respect to the laboratory frame, the four-momentum transforms as
\begin{align}
    E_{\rm CNB}=\gamma_w (E_{\rm LAB}-p_{\rm LAB}v_w u)\,,\quad p_{\rm CNB(\parallel)} = \gamma_w(p_{\rm LAB(\parallel)}-E_{\rm LAB}v_w)\,,\quad p_{\rm CNB(\perp)}=p_{\rm LAB(\perp)}\,,
    \label{eq:Lorentz boost from LAB to CNB}
\end{align}
with $\gamma_w=(1-v_w^{2})^{-1/2}$ and $u\equiv \hat{\bm p}_{\rm LAB}\cdot \hat{\bm v}_w=p_{\rm LAB(\parallel)}/p_{\rm LAB}$ where $\parallel$ and $\perp$ denote components parallel and perpendicular to $\bm v_w$, respectively.

We define
\begin{align}
     \bar{v}_{E}\equiv \frac{T_{\nu}}{E_{\rm LAB}}\,,\quad 
     x\equiv \frac{p_{\rm LAB}}{T_{\nu}}\,,\quad v_\nu\equiv\frac{p_{\rm LAB}}{E_{\rm LAB}}=x\bar{v}_{E}\,,
\end{align}
Using this definition, the magnitude of the CNB-frame momentum can be written as
\begin{align}
    p_{\rm CNB}=\gamma_w T_\nu \sqrt{x^{2} - 2 x \frac{v_w}{\bar{v}_{E}}\cos\theta_{\nu w} + \left(\frac{v_w}{\bar{v}_{E}}\right)^{2}- x^{2}v_w^{2}\sin^{2}\theta_{\nu w}}\,.
\end{align}
In the limit $v_w\ll1$, while keeping $v_w/\bar v_E$ arbitrary, this reduces to
\begin{align}
    p_{\rm CNB}\simeq T_\nu \sqrt{x^{2} - 2 x \frac{v_w}{\bar{v}_{E}}\cos\theta_{\nu w} + \left(\frac{v_w}{\bar{v}_{E}}\right)^{2}}=|\bm{p}_{\rm LAB}-E_{\rm LAB}\bm{v}_w|\,.\label{eq:pCNB in v_w ll 1 limit}
\end{align}
If we further assume $p_{\rm LAB}\ll m_\nu$, \eqref{eq:pCNB in v_w ll 1 limit} reduces to
\begin{align}
    p_{\rm CNB}= |\bm{p}_{\rm LAB}-m_\nu \bm{v}_{w}|\left(1+O\left(\frac{p_{\rm LAB}^{2}}{m_\nu^{2}}\right)\right)\,,\quad \text{with }\bar{v}_{E}\simeq \bar{v}_{\nu}\equiv  \frac{T_{\nu}}{m_\nu}\,.
\end{align}
On the other hand, if we further take $v_w/\bar{v}_{E}\ll 1$, 
\eqref{eq:pCNB in v_w ll 1 limit} reduces to
\begin{align}
    p_{\rm CNB}= p_{\rm LAB}-T_\nu\frac{v_{w}}{\bar{v}_{E}} \cos{\theta_{w \nu}} + O\left(T_\nu\frac{v_{w}^{2}}{\bar{v}_{E}^{2}}\right)\,,
\end{align}
corresponding to the dipole approximation of the distribution,
\begin{align}
    f_{\rm LAB}(p_{\rm LAB}) =& f_{\rm CNB}(p_{\rm CNB})=f_0(p_{\rm CNB})\nonumber\\
    =& f_0(p_{\rm LAB}) +f_1(p_{\rm LAB})\cos{\theta_{\nu w}}+O\left(\frac{v_{w}^{2}}{\bar{v}_{E}^{2}}\right)\,,
\end{align}
where
\begin{align}
    f_0(p)=\frac{1}{\exp(p/T_\nu)+1}\,,\quad f_1(p)=f_0(p)(1-f_0(p))\frac{v_w}{\bar{v}_{E}}\,.
\end{align}
The two approximations used above, \(p_{\rm LAB}\ll m_\nu\) and \(v_w\ll \bar v_E\), are independent limits. They may overlap in part of parameter space, but neither implies the other. As discussed in the main text, the latter condition is satisfied only for sufficiently light neutrinos, roughly \(m_\nu \ll 0.1\,{\rm eV}\).
The approximate CNB flux in \eqref{eq:dipole approximation of flux} is valid in the range both \(p_{\rm LAB}\ll m_\nu\) and \(v_w\ll \bar v_E\simeq \bar v_\nu\) are valid.

We neglect the decoupling-era mass correction $O(m_\nu^2/T_{\rm dec}^2)$. It is distinct from the mass dependence generated by the Lorentz boost: since the boost mixes momentum and energy, the dipole term contains $E_\nu/T_\nu$, which becomes $m_\nu/T_\nu$ for nonrelativistic neutrinos.
This behavior is consistent with the discussion around (3.11) of Ref.~\cite{Shergold:2021evs}, where the Earth-frame momentum of unclustered relic neutrinos is shown to be controlled either by the thermal momentum or by the boost momentum, depending on the relative size of the thermal velocity and the Earth–CNB relative velocity. In the present calculation the same boost effect appears in the dipole correction to the distribution function: the Lorentz transformation mixes momentum and energy, so that the dipole term is proportional to $E_\nu/T_\nu$, which becomes $m_\nu/T_\nu$ for nonrelativistic neutrinos.

\section{Derivation of detection rate}
\label{sec:detection rate}

In this appendix we derive \eqref{eq:dGamma_CNB} starting from \eqref{eq:dGamma}. To simplify the notation we omit the $X$ index to distinguish the matrix element in the Dirac and Majorana cases.
We decompose the distribution function with the Legendre polynomial,
\begin{align}
    f_\nu(p_\nu,u_{\nu w})=\sum_{\ell=0}^{\infty}{f}_\ell(p_\nu)P_{\ell}(u_{\nu w})\,.
\end{align}
Inserting it together with \eqref{eq:matrix element} into \eqref{eq:dGamma}, we obtain
\begin{align}
    \frac{\D \Gamma_{\rm CNB}^{X}}{\D \Omega_e}
    =&
    \frac{G_{F}^{2}}
    {2(2\pi)^{9}}|U_{ei}|^{2}|V_{ud}|^{2} \sum_{\ell}\int_0^{\infty}\D p_\nu p_\nu^{2}f_\ell(p_\nu)\int_0^{\infty}\D p_{\He} p_{\He} ^{2}\int_0^{\infty}\D p_e p_e^{2}\nonumber\\
    &\times F(Z,E_e)(2\pi)\delta (E_\nu + m_{\Tritium}-E_{\He}-E_{e})
    [C_{A}A\big\langle P_{\ell}(u_{\nu w})\big\rangle+C_{B}Bv_e\big\langle u_{\nu e}P_{\ell}(u_{\nu w})\big\rangle]\,,
\end{align}
where $\langle\cdots\rangle$ denotes the angular integral with the definition,
\begin{align}
    \langle q\rangle \equiv \int \D \Omega_\nu\int \D \Omega_{\He}(2\pi)^{3}\delta^{(3)}(\bm{p}_\nu -\bm{p}_{\He}-\bm{p}_{e}) \,q\,.
\end{align}

The angular integration over $\Omega_{\rm He}$ is performed using two components of the three-dimensional delta function. The remaining constraint fixes the angle between $\bm p_\nu$ and $\bm p_e$ as 
\begin{align}
    u_{\nu e}=u_*\equiv \frac{p_\nu^{2}+p_e^{2}-p_{\He}^{2}}{2p_\nu p_e}\,,
\end{align}
which follows from $p_{\He}=|\bm{p}_{\nu}-\bm{p}_{e}|$. The corresponding support condition $|u_\ast|\leq 1$ is equivalent to
\begin{align}
    \Theta_\Delta \equiv \Theta(p_{\He}-|p_\nu - p_e|)\Theta(-p_{\He} + p_\nu+p_e)\,.
\end{align}
The remaining nontrivial angular integral is the azimuthal integral around $\bm{n}_{e}$. Since $u_{\nu w}=u_{\nu e}u_{ew}+\cos(\phi_{\nu e}-\phi_{ew})\sqrt{1-u_{\nu e}^2}\sqrt{1-u_{ew}^2}$, the azimuthal average of $P_\ell(u_{\nu w})$ gives
\begin{align}
    \frac{1}{2\pi}\int_{0}^{2\pi}\D\phi_{\nu e}P_{\ell}(u_{\nu w}) = P_{\ell}(u_{\ast})P_\ell(u_{ew})\,.
\end{align}
Thus we obtain
\begin{align}
    \frac{\D \Gamma_{\rm CNB}^{X}}{\D \Omega_e}
    =&
    \frac{G_{F}^{2}}
    {2(2\pi)^{5}}|U_{ei}|^{2}|V_{ud}|^{2}\sum_{\ell=0}^{\infty} P_{\ell}(u_{ew})\int_0^{\infty}\D p_\nu p_\nu {f}_\ell(p_\nu)\int_0^{\infty}\D p_{\He} p_{\He} \int_0^{\infty}\D p_e p_e\nonumber\\
    &\times F(Z,E_e)(2\pi)\delta (E_\nu + m_{\Tritium}-E_{\He}-E_{e})
    \Theta_{\Delta}[C_{A}A+C_{B}Bv_eu_*]P_{\ell}(u_*)\,.
\end{align}
Note that $u_{ew}=\bm{n}_{e}\cdot\bm{n}_{w}$ is constant over these integrals and can be factored out.

\paragraph{Radial momentum integrations:}

Since both the energy-conservation Dirac delta function and $\Theta_\Delta$ are functions of $(p_e,p_{^3{\rm He}})$, performing these integrals exactly is technically cumbersome. 
For example, after integrating over $p_e$ with the Dirac delta function associated with energy conservation, the integration boundaries for $p_{\He}$, $[|p_e(p_{\He},p_\nu)-p_\nu|,\,\,p_e(p_{\He},p_\nu)+p_\nu]$ becomes $p_{\He}$-dependent. Here $p_e(p_{\He},p_\nu)=\sqrt{E_e^{2}(p_{\He},p_\nu)-m_e^{2}}$ and $E_{e}(p_{\He},p_\nu)=m_{\Tritium}+E_\nu-E_{\He}$.

However, the width of the range $\Delta p_{\He}|_{{\rm fixed} p_\nu}\sim 2p_\nu$ is much narrower than $O(p_e(p_{\He},p_\nu))$ if $p_\nu\ll p_e(p_{\He},p_\nu)$. 
Integrating over $\D p_e$, we obtain
\begin{align}
    \frac{\D \Gamma_{\rm CNB}^{X}}{\D \Omega_e}
    =&
    \frac{G_{F}^{2}}
    {2(2\pi)^{4}}|U_{ei}|^{2}|V_{ud}|^{2}\sum_{\ell=0}^{\infty} P_{\ell}(u_{ew})\int_0^{\infty}\D p_\nu p_\nu f_\ell(p_\nu)\int_0^{\infty}\D p_{\He} p_{\He} E_{e}(p_{\He},p_\nu)\nonumber\\
    &\times F(Z,E_e(p_{\He},p_\nu))
    \Theta_{\Delta}[C_{A}A+C_{B}Bv_eu_*]P_{\ell}(u_*)\,.
\end{align}
As we will show later, one can approximate $E_{e}F(Z,E_{e})$ as a constant in the narrow range of the integral over $p_{\He}$, and obtain
\begin{align}
    \frac{\D \Gamma_{\rm CNB}^{X}}{\D \Omega_e}
    \simeq &
    \frac{G_{F}^{2}}
    {2(2\pi)^{4}}|U_{ei}|^{2}|V_{ud}|^{2}\sum_{\ell=0}^{\infty} P_{\ell}(u_{ew})\int_0^{\infty}\D p_\nu p_\nu f_\ell(p_\nu)
    F(Z,E_{*})E_{*}
    \nonumber\\
    &\times \int_{p_*-p_\nu}^{p_*+p_\nu}\D p_{\He} p_{\He} [C_{A}A+C_{B}Bv_eu_*]P_{\ell}(u_*)\,,
\end{align}
where the starred quantities are the two-body kinematic solutions defined in \eqref{eq:starred Ee}.
We further approximate $\D p_{^3{\rm He}}\,p_{^3{\rm He}}\simeq p_\ast p_\nu\,\D u_\ast$ where the recoil correction to this change of variables is of order $O(E_\ast/E_{^3{\rm He}})\sim 10^{-4}$. 
Using the orthogonality of the Legendre functions, 
\begin{align}
    \int_{-1}^{1}\D u P_{\ell}(u)P_{n}(u) = \frac{2}{2\ell + 1}\delta_{n\ell}\,,
\end{align}
we have
\begin{align}
    \frac{\D \Gamma_{\rm CNB}^{X}}{\D \Omega_e}
    \simeq &
    \frac{G_{F}^{2}}
    {(2\pi)^{4}}|U_{ei}|^{2}|V_{ud}|^{2}\sum_{\ell=0}^{\infty} P_{\ell}(u_{ew})\int_0^{\infty}\D p_\nu p_\nu^{2} f_\ell(p_\nu)
    F(Z,E_{*})E_{*}p_*\left[C_{A}A\delta_{\ell 0}+\frac{1}{3}C_{B}Bv_e\delta_{\ell 1}\right]\,.
\end{align}
Using $v_e=p_*/E_*$, we obtain
\begin{align}
    \frac{\D \Gamma_{\rm CNB}^{X}}{\D \Omega_e}
    \simeq &
    \frac{G_{F}^{2}}
    {(2\pi)^{4}}|U_{ei}|^{2}|V_{ud}|^{2}\int_0^{\infty}\D p_\nu p_\nu^{2} 
    F(Z,E_{*})\left[C_{A}AE_{*}p_*f_0(p_\nu)+\frac{1}{3}C_{B}Bp_*^{2}f_1(p_\nu)u_{ew}\right]\,\nonumber \\
    =&
    \frac{G_{F}^{2}}
    {(2\pi)^{4}}|U_{ei}|^{2}|V_{ud}|^{2}\left[C_{A}\mathcal{I}_{A}+ \frac{1}{3}C_{B}\mathcal{I}_{B}u_{ew}\right]\,,
\end{align}
where the definitions of $\mathcal{I}_{A}$ and $\mathcal{I}_{B}$ are given in \eqref{eq:A,B for M,D}. 
This approximate expression agrees with the direct numerical integration up to recoil-suppressed corrections of order $O(E_e/m_{^3{\rm He}})\sim 10^{-4}$.  This correction is larger than the effect of the finite width of the $p_{^3{\rm He}}$ integration region, which is only $O(p_\nu/p_e)\sim 10^{-9}$.

\paragraph{Justification of  the constant $E_{e}F(Z,E_{e})$ approximation in the narrow $p_{\He}$-integrals:}
Below we show
\begin{align}
    \left|\frac{\Delta [E_{e}F(Z,E_{e})]/[E_{e}F(Z,E_{e})]}{\Delta [p_{\He} P_{\ell}(u_*)]/[p_{\He} P_{\ell}(u_*)]}\right|\ll1\,,\quad
    \left|\frac{\Delta [E_{e}F(Z,E_{e})]/[E_{e}F(Z,E_{e})]}{\Delta [p_{\He} u_*P_{\ell}(u_*)]/[p_{\He} u_*P_{\ell}(u_*)]}\right|\ll1\,,
    \label{eq: condition to ignore the change of EeF}
\end{align}
and therefore we can treat $E_{e}F(Z,E_{e})$ as a constant when we perform $p_{\He}$-integrals. Here $\Delta[X]=X|_{p_{\He}=p_e+p_\nu} - X|_{p_{\He}=p_e-p_\nu}$.
One can confirm this as follows. 
With the change of $\Delta p_{\He}=2p_\nu\ll p_e$, we have
\begin{align}
    \frac{\Delta p_{\He}}{p_{\He}} = \frac{2p_\nu}{p_{e}}\,,\quad 
    \frac{\Delta u_*}{u_*} = - \frac{4p_e}{p_\nu}\,,\quad 
    \frac{\Delta E_{e}}{E_{e}} = - \frac{2p_\nu}{E_{e}}\frac{p_e}{E_{\He}}\simeq - \frac{2p_\nu}{E_{e}}\frac{p_e}{m_{\He}}\,,
\end{align}
where we took the $p_{\He}=p_e$ as the fiducial value out of $\Delta[\cdots]$ and used $E_{\He}\simeq m_{\He}\gg p_e$.
In the same limit $p_\nu\ll p_e$ we also have
\begin{align}
    \frac{\D \ln P_{\ell}(u_*)}{\D u_*}\simeq \begin{cases}
        0\,,&\quad\ell=0\,,\\
        O\left(\frac{1}{u_*}\right)=O\left(\frac{p_e}{p_\nu}\right)\,,&\quad \ell=1,3,\cdots\\
        O(\ell^{2}u_*)=O\left(\ell^{2}\frac{p_\nu}{p_e}\right)\,,&\quad \ell=2,4,\cdots
    \end{cases}
\end{align}
for a fixed $\ell$ and $O(u_{*})=O( p_\nu/p_e)\ll 1$.
Therefore,
\begin{align}
    \frac{\Delta [p_{\He} P_{\ell}(u_*)]}{p_{\He} P_{\ell}(u_*)} =& 2\left(\frac{p_\nu}{p_e}-\frac{\D \ln P_{\ell}(u_*)}{\D u_*}\right)\simeq \begin{cases}
        O(u_*)\,,&\quad\ell=0\,,\\
        O\left(\frac{1}{u_*}\right)\,,&\quad \ell=1,3,\cdots\\
        O(\ell^{2}u_*)\,,&\quad \ell=2,4,\cdots\\
    \end{cases}
    \label{eq:variation of pHe Pell}
\end{align}
and
\begin{align}
    \frac{\Delta [p_{\He} u_*P_{\ell}(u_*)]}{p_{\He} u_*P_{\ell}(u_*)} =& 2\left(\frac{p_\nu}{p_e}-\frac{\D \ln P_{\ell}(u_*)}{\D u_*}-\frac{1}{u_*}\right)\simeq O\left(\frac{1}{u_*}\right)\,\quad \forall\ell\,.
    \label{eq:variation of pHe uPell}
\end{align}

On the other hand, we have
\begin{align}
    \frac{\Delta [E_{e}F(Z,E_{e})]}{E_{e}F(Z,E_{e})} = \frac{\Delta E_{e}}{E_{e}}\left(1+\frac{d\ln F}{d\ln E}\right) = \frac{\Delta E_{e}}{E_{e}}\left(1+(e^{-2\pi\eta}F-1)\frac{m_e^2}{p_e^2}\right)\,.
\end{align}
Using $p_e/m_e\simeq v_e\simeq O(0.1)$ and $|1-e^{-2\pi\eta}F|\sim O(1)$, we obtain
\begin{align}
    \left|\frac{\Delta [E_{e}F(Z,E_{e})]}{E_{e}F(Z,E_{e})}\right|\simeq \frac{\Delta E_{e}}{E_{e}} \frac{m_e^2}{p_e^2}\,.
\end{align}
By comparing it with \eqref{eq:variation of pHe Pell} and \eqref{eq:variation of pHe uPell}, we can explicitly confirm \eqref{eq: condition to ignore the change of EeF} as 
\begin{align}
\begin{split}
    \left|\frac{\Delta [E_{e}F(Z,E_{e})]/[E_{e}F(Z,E_{e})]}{\Delta [p_{\He} u_*P_{\ell}(u_*)]/[p_{\He} u_*P_{\ell}(u_*)]}\right|
    \lesssim&
    \left|\frac{\Delta [E_{e}F(Z,E_{e})]/[E_{e}F(Z,E_{e})]}{\Delta [p_{\He} P_{\ell}(u_*)]/[p_{\He} P_{\ell}(u_*)]}\right|\\
    \lesssim& \left(\frac{p_e}{p_\nu}\right)\left(\frac{p_\nu}{E_{e}}\frac{p_e}{m_{\He}}\frac{m_e^2}{p_e^2}\right)
    \\\simeq &\frac{m_e}{m_{\He}}\ll 1\,.
\end{split}
\end{align}

\bibliography{reference} 
\bibliographystyle{JHEP}

\end{document}